\documentclass{aa}  

\usepackage{graphicx}
\usepackage{adjustbox}
\usepackage{txfonts}
\newcommand{\cifig}[1]{Figure~\ref{#1}} 
\begin{document}

   \title{Dust distribution in circumstellar disks harboring multi-planet systems}

   \subtitle{II. Super-thermal mass planets}

   \author{V. Roatti
          \inst{1}
          \and
          G. Picogna\inst{2}
          \and
          F. Marzari\inst{3}
          }

   \institute{Dipartimento di Fisica e Astronomia “G. Galilei”, Universit\`a degli Studi di Padova, Vicolo dell’Osservatorio 3, 35122 Padova, Italy\\
              \email{vincenzo.roatti@phd.unipd.it}
         \and
             Universit\"ats-Sternwarte, Ludwig-Maximilians-Universit\"at M\"unchen, Scheinerstr. 1, M\"unchen, 81679, Bayern, Germany\\
             \email{picogna@usm.lmu.de}
\and
Dipartimento di Fisica e Astronomia “G.Galilei”, Universit\`a degli Studi di Padova, Via Marzolo 8, 35121 Padova, Italy \\
\email{francesco.marzari@pd.infn.it}
             }

   \date{Received ... / Accepted ...}
 
  \abstract 
   {
   Theoretical formation models and exoplanet detection surveys indicate that systems with multiple giant planets are common.
 }
   {We investigate how multiple super-thermal mass planets embedded in a circumstellar disk shape the dust distribution and examine the consequences for interpreting disk substructures and inferring planetary properties.}
   {We perform two-dimensional hydrodynamical simulations with a modified PLUTO code, treating dust as Lagrangian particles in a wide range of sizes. 
   We analyze systems with two planets of different masses and orbital separations, comparing them to the single-planet scenario.
We generate synthetic ALMA continuum maps using RADMC-3D and compute the relative impact velocities of dust particles to assess potential limitations to grain growth.}
   {
   Dust morphologies in multi-planet systems cannot be described as a simple superposition of single-planet gaps. Secular planetary perturbations can generate multiple dust traps and asymmetric structures, while also exciting significant eccentricities in dust particle orbits. As a consequence, the locations and widths of dust rings and gaps depend on the size of the particles, the masses of the planet, and the orbital configurations. Synthetic continuum images may hide gaps carved by multiple planets, thereby complicating the interpretation of observed substructures. {In addition, eccentricities induced in dust orbits lead to stronger gas drag, reducing the Stokes number for a given particle size,} and the enhanced relative velocities associated with eccentric orbits can further suppress grain growth, promoting fragmentation and replenishment of small dust grains.
   }
   {}


   \maketitle

\section{Introduction}

High-angular-resolution observations of millimeter dust continuum emission in circumstellar and protoplanetary disks have profoundly advanced our understanding of disk structure and evolution \citep{ALMA2015, avenhaus_sphere_2018, 2018ApJ...869L..41A}. These observations reveal a remarkable diversity in disk sizes and morphologies and demonstrate that disks are commonly characterized by substructures such as rings, gaps, spirals, and asymmetries \citep{isella2018ApJ}. Among these features, gap–ring systems are ubiquitous and appear to be a fundamental outcome of disk evolution rather than exceptional configurations. \\*
Several physical mechanisms have been proposed to account for the origin of these substructures. One prominent explanation involves planet–disk interactions, in which embedded planets gravitationally perturb the disk and carve gaps in the gas and dust components \citep{crida2006, dong_fung2017}. Alternative scenarios include dust growth and fragmentation near snow lines, where changes in grain composition enhance dust accumulation \citep{zhang2015}, as well as magneto-hydrodynamic processes such as turbulence driven by magneto-rotational instability \citep{flock2015}, MHD winds \citep{riols_MHD}, and pressure maxima associated with transitions between magnetically active and dead zones \citep{dzyu2010}. Distinguishing between these mechanisms remains a key challenge in interpreting high-resolution observations. \\*
In this work, we focus on the formation of gaps and rings induced by giant planets embedded in circumstellar disks. In \citet{2025A&A...703A.270R}, we studied sub-thermal mass planets, where the thermal mass, M$_{\rm{th}}$, is defined as the planetary mass at which the Hill radius of the planet, $r_H$, is equal to the vertical scale height of the disk: $(M_{p} / M_{\star})_{th} = 3(H/r _{p})^3$, with $r_p$ the radius of the planet's orbit. In that case, we illustrated how the planets were able to open dust gaps without perturbing the gas distribution. In general, a single sufficiently massive planet can open a gap in the gas disk through tidal torques, provided that its gravitational perturbation exceeds the stabilizing effects of viscosity and pressure \citep{linpapa1986}. The resulting gap is accompanied by the formation of pressure maxima at its edges, which act as efficient dust traps by halting the inward or outward radial drift of solid particles \citep{weidenschilling1977, rosotti2020}. Dust initially residing within the gap is rapidly depleted and redistributed toward these pressure maxima, where it accumulates and forms ring-like structures. In particular, dust drifting inward from the outer disk tends to pile up efficiently at the outer edge of the gap, leading to a pronounced dust ring that is frequently observed in disks hosting giant planets \citep{pinilla2012, isella2016}. \\*
If the planet migrates radially, the structure and efficiency of the associated dust traps evolve over time. The morphology and relative brightness of the resulting dust rings depend sensitively on both the direction and the rate of migration, introducing an additional layer of complexity when interpreting the observed disk substructures. \\*
The problem becomes substantially more complex in systems hosting multiple planets, particularly giant ones. Formation models (e.g., \cite{ida2013};
 \cite{matsumura2021}) and radial velocity surveys
(\cite{knutson2014}; \cite{zhu2022})  suggest that
multiple giant planets often form and coexist. The combined gravitational influence of two or more planets can significantly alter the gas surface density profile and, consequently, the efficiency and location of the dust traps associated with each planet. In such cases, the resulting dust distribution may differ markedly from that expected from a simple superposition of the individual gaps and rings produced by isolated planets. This has important implications for attempts to infer planetary masses and orbital configurations from observed dust substructures, as assuming that each planet acts independently may lead to biased or inaccurate conclusions. \\*
Previous studies have explored some aspects of this regime. In particular, \cite{marzari_angelo2019} investigated the case of two resonant giant planets that open a common gap in gas and dust and, while migrating outward, induce a strong dust accumulation at the outer edge of the gap. Here, we instead focus on systems hosting two widely separated giant planets, a configuration that is particularly relevant for disks exhibiting multiple, well-separated rings. \\*
We perform hydrodynamical simulations using the PLUTO code \citep{mignone2007}, modified to include Lagrangian dust particles spanning a broad range of sizes, as in \citet{2018A&A...616A.116P}. To provide a more realistic description of dust evolution, we implement a constant particle flux at the outer disk boundary, include diffusive processes, and account for dust sublimation at the water snow line. These ingredients allow us to follow the coupled evolution of gas and dust under the combined influence of multiple planets. \\*
We find that the dust structures produced by multiple, widely separated giant planets cannot, in general, be interpreted as a linear superposition of the gaps and rings generated by each planet individually. We show that the combined gravitational perturbations of the planets lead to complex gas pressure profiles that give rise to non-trivial dust trapping structures and induce significant eccentricities in the dust particle orbits. As a result, the dust morphology exhibits features that are qualitatively different from those expected from single-planet models, even when the planets are well separated. \\*
In addition, we post-process the hydrodynamical simulations using the radiative transfer code RADMC-3D \citep{2012ascl.soft02015D} to generate synthetic dust continuum emission maps in several ALMA observing bands. This allows us to assess whether the physical properties of the planetary system, most notably the planetary masses and orbital parameters, can be reliably recovered from continuum observations. In particular, we investigate the extent to which observed disk substructures can be misinterpreted when the presence of multiple planets is not properly accounted for. For this purpose, we feed our synthetic images to DBNets2.0 \citep{2025A&A...700A.190R}, a machine learning tool that observers can use to estimate the mass of a single embedded planet provided the location of the observed gap. Therefore, we can find the mismatch between the real and derived mass in the case of a multi-planet system. \\*
Finally, we compute the relative impact velocities between dust particles to evaluate whether the eccentricities induced by the planets lead to enhanced collision velocities that may inhibit further grain growth. This provides additional insight into the interplay between the disk substructure, dust dynamics, and the early stages of planet formation. \\*
This paper is organized as follows. In Sect. 2, we describe the modifications to the PLUTO code that enable a more realistic treatment of dust evolution. In Sect. 3, we present the initial setup of the simulations. In Sect. 4, we validate the model by considering the case of a single giant planet. In Sect. 5, we examine systems with two equal-mass planets in different orbital configurations, including both closely spaced and widely separated systems. In Sect. 6, we extend the analysis to planets of unequal masses, considering a Jupiter–Saturn pair. In Sect. 7, we present the synthetic continuum images obtained with RADMC-3D. In Sect. 8, we analyze the impact velocities between dust particles and discuss their implications for dust growth. Finally, in Sect. 9, we summarize our results and discuss their broader implications.

\section{The code PLUTO, additional features}

PLUTO \citep{mignone2007} is a grid-based versatile hydrodynamics and magnetohydrodynamics code designed to solve problems of astrophysical fluid dynamics with high precision.  It is well-suited for modeling the complex gas and dust dynamics in circumstellar disks. To make the simulations more realistic, we implemented several features (see \citet{2025A&A...703A.270R} for more details on the numerical setup). First, we considered a thin disk without self-gravity, which can be described as either globally or locally isothermal. To handle the dynamics of the planets, we added an N-body module as in \citet{2018A&A...616A..47T}. Instead of introducing the planets right at the beginning of the simulation, we gradually increased their mass from 0 to their final mass with either a linear or exponential growth profile, to avoid significant perturbations of the gas and particles \citep{2006MNRAS.370..529D}. \\*
We computed the acceleration due to the mutual gravitational attraction between the planets and the star as in \cite{2018A&A...616A..47T}. We smoothed the gravitational potential to avoid singularities in the numerical evaluation of the acceleration: in particular, we used a smoothing length of $\epsilon = 0.6$ H, as this value describes the vertically averaged forces very well \citep{2012A&A...541A.123M, 2025MNRAS.543.4198C}. Moreover, we calculated the gravitational feedback of the disk onto the planets (and the star) to account for planet migration \citep{2018A&A...616A..47T}. \\*
We modeled the solid component of the disk using a large number of Lagrangian particles representative of dust dynamics \citep{2018A&A...616A.116P}. We chose this approach over the pressure-less fluid to be able to follow individual particles' trajectories. Moreover, the fluid approach is not particularly accurate when working with large particle sizes. We {initialized $N = 5 \times 10^5$ particles with} a multi-size distribution divided into 10 bins from 100 $\mu$m to 5.12 cm, separated by powers of 2. Particle orbits are numerically integrated with a semi-implicit scheme \citep{2014ApJ...785..122Z} under the gravitational influence of the massive bodies, while also experiencing local gas drag, which is computed by interpolating the Epstein and Stokes regimes as in \citet{2009A&A...493.1125L}:
\begin{equation} \label{Cd}
    C_D = \frac{9\mathrm{Kn}^2 C _D ^{\mathrm{Eps}} +C _D ^{\mathrm{Stk}} }{(3\mathrm{Kn} + 1)^2}\,,
\end{equation}
where $C _D ^{\mathrm{Eps}}$ and $C _D ^{\mathrm{Stk}}$ are the Epstein and Stokes drag coefficients, respectively, and Kn = $\lambda / 2s$ is the Knudsen number, with $\lambda$ the mean free path of the gas and $s$ the particle size \citep{2003A&A...399..297W, 2009A&A...493.1125L, 2018A&A...616A.116P}. From equation \ref{Cd} we can compute the dust stopping time,
\begin{equation} \label{taus}
    \tau _s = \frac{4 \lambda \rho _d}{3 \rho _g C_D c_s}\frac{1}{\mathrm{MaKn}}\,,
\end{equation}
where $\rho _d$ = 1 g/cm$^3$ is the internal density of dust particles, $\rho _g$ is the density of gas, $c_s$ is the speed of sound, and $\rm{Ma} = |{v}_{\rm{rel}}|/c_s $ is the Mach number. Then, the drag force acting on a dust grain moving with velocity $\mathbf{v}_{\mathrm{rel}}$ relative to the gas is given by
\begin{equation}
    F_D = - \frac{\mathbf{v}_{\mathrm{rel}}}{\tau _s} .
\end{equation}\\*
Then we calculate the location of the water snowline as the radius at which the pressure of the water vapor is equal to the equilibrium pressure \citep[see][eqs. 33 and 34]{garate2020}. {The equations define the thermodynamic threshold for the phase transition. The abundance of water is set by the initial solid-to-gas ratio. Following \cite{2017A&A...608A..92D}, we assumed an initial global dust-to-gas ratio of 0.01. Beyond the snowline, water ice typically constitutes about 50\% of the total solid mass (or a water-to-gas mass ratio of roughly $5 \times 10^{-3}$)}. At the snowline, the sizes of particles crossing the line are reduced by a factor of two due to ice sublimation. As a result, particles transition to a smaller size bin. {Because we assumed the same gas and temperature profiles in all of our models, the water snowline is always located at $r_{\rm{snow}} \simeq$ 3 au.} \\*
Moreover, a constant flux of dust from the outer disk regions is simulated by maintaining a steady number of dust particles in the outer ring of the grid {, replacing the particles lost from the inner boundary because of dust drift}. This allows the numerical integration to continue over a long period of time without the inconvenience of exhausting the flux of dust particles from the outer regions of the disk. Note that this makes sense physically as a real disk extends further than the simulation domain. {For the gas component, we added wave-killing zones at the inner and outer radial boundaries of the grid, so that the surface density of the gas is damped towards its initial value \citep{2006MNRAS.370..529D}. However, the integration time is too short for the viscous evolution of the disk to have a noticeable effect. Specifically, for a disk with characteristic size $r$, the surface density at all radii will evolve on a time scale $\tau_\nu = r^2/\nu$ \citep{2020apfs.book.....A}. For our disk parameters this translates into a viscous time-scale of $6.4 \times 10^6$ yr, which is more than two orders of magnitude larger than our integration time of 270 orbits at 22 au: $t = 2.8 \times 10 ^4$ yr. Our constant disk surface density at the outer boundary provides an effective gas influx from the outer disk, which is supposed to extend much further than our computational domain and matches the constant particle surface density at the outer boundary.} \\*
Finally, we included the diffusion of dust particles following \cite{charnoz2011}, incorporating periodic position jumps to simulate turbulent mixing.

\subsection{Postprocessing with RADMC3D}
An important aspect of modeling dust evolution in protoplanetary disks hosting planets is the interpretation of observed substructures such as rings, gaps, and spiral arms. In order to achieve this goal, our simulations include a broad range of dust grain sizes, which is essential for accurately reproducing ALMA observations at different wavelengths and frequency channels. By capturing how dust of different sizes responds to the presence of planets, we aim to establish a robust link between disk morphology and planetary perturbations. Ultimately, the main objective is to tackle the inverse problem, namely to infer the physical properties of the embedded planets, such as their masses and orbital parameters, from the observed disk features.
With this goal in mind, we post-processed the output of the hydrodynamical simulations using the radiative transfer code RADMC-3D \citep{2012ascl.soft02015D} to generate synthetic dust continuum emission maps at wavelengths of 0.85, 1.3 and 3 mm, corresponding to ALMA Bands 7, 6, and 3, respectively. Dust density distributions were calculated using a modified version of the FARGO2RADMC3D code \citep{2019MNRAS.486..304B}. We assumed a standard power-law dust size distribution $n(a) \propto a^{-3.5}$, covering the minimum and maximum grain sizes considered in the simulations. Since dust back-reaction on the gas was neglected, the dust mass does not affect the gas dynamics; consequently, only the spatial distribution of the dust enters the continuum radiative transfer calculations \citep{2019MNRAS.486..304B}. \\*
The dust surface density for each size bin was calculated from its spatial distribution as follows \cite{2019MNRAS.486..304B}:
\begin{equation}
  \sigma _{i,\rm{dust}}(r, \theta) = \frac{N_i (r,\theta)}{A(r)}\frac{M_{i,\rm{dust}}}{\sum _{r,\theta}N_i (r,\theta)}, 
\end{equation}
where $N_i$ is the number of particles per bin size in each grid cell, $A(r)$ is the cell surface area (seen face-on) and
\begin{equation}
    M_{i,\rm{dust}} = \xi M_{\rm{gas}}\frac{a_i ^{4-p}}{\sum _i a_i ^{4-p}}
\end{equation}
is the mass of dust per size bin. Here, $p = 3.5$ is the power-law index of the size distribution, $\xi = 0.01$ the dust-to-gas ratio, and $M_{\rm{gas}}$ the total gas mass.
To make the synthetic images more realistic, we included an additional dust bin with small particles (1 $\mu$m) that are tightly coupled to the gas. \\*
We expanded the dust surface density along the vertical direction (128 cells from 0 to $\pi$) to obtain the dust volume density. We assumed hydrostatic equilibrium and a Gaussian vertical profile for each size bin, where the dust particle scale height for the $i$-th size bin was set as $H_{i, \rm{dust}}/H = \sqrt{\alpha/(\alpha + \rm{St _i})}$ \citep{2007Icar..192..588Y}, where St = $\tau_s \Omega _K$ is the Stokes number of the particles, with $\tau _s$ the stopping time and $\Omega _K$ the Keplerian frequency. Dust opacities were calculated using the \textit{optool} code \citep{2021ascl.soft04010D}, which adopts the Mie theory for compact spherical grains with the optical constants of astronomical silicates \citep{2003ARA&A..41..241D}. \\*
For the synthetic observations, we placed the disk at a distance of 140 pc, which is typical of close star-forming regions, with the central star of radius $R = 2.4$ $R_{\odot}$ and temperature $T = 4300$ K. The dust temperatures were computed with a Monte Carlo calculation. We adopted a logarithmically-spaced wavelength grid with 300 elements $5 \times 10^{-2}\leq \lambda [\mu m] \leq 10^4$ and we used $10^8$ photon packages for ray-tracing, generating 2048 $\times$ 2048 pixel flux maps which include both absorption and scattering, assuming Henyey-Greenstein anisotropic scattering. \\*
\section{Initial set-up}
\begin{table}
\caption{Initial orbital configurations of the simulated models.}
\label{tab:orb}
\centering
\begin{tabular}{ c  c  c  c  c  }
\hline\hline
{Model} & {M$_{1}$ (M$_{j}$)} & {M$_{2}$ (M$_{j}$)} & {a$_1$ (au)} & {a$_2$ (au)}   \\
\hline
Single &    1 & / & 5 & / \\
CloseJup & 1 & 1 & 8 & 15 \\
WideJup & 1          & 1   & 5  & 22 \\
WideSat & 1          & 0.3   & 5 & 22 \\
\hline
\end{tabular}
\end{table}
We perform two-dimensional simulations of a circumstellar disk in which two giant planets have formed at a significant distance from each other such that there is no resonant locking between them. {The initial orbital configuration for each model is shown in Table \ref{tab:orb}.} Since the planets are assumed to have completed their formation, the surface density of the gaseous disk is set to a low value, with its profile described by a power law of the form:
\begin{equation}
    \Sigma = \Sigma_0 r^{-1} ,
\end{equation}
{with the initial density $\Sigma_0 = 100$ g/cm$^2$ in all models}, as we are considering aged disks in which planet formation has already been completed. The temperature profile is given by 
\begin{equation}
    T(r) = T_0 r^{-0.5}.
\end{equation}
The initial temperature at 1 au is set in all models to $T_0 = 300 $ $K$, while an 
alpha--viscosity prescription is adopted \citep{1973A&A....24..337S}, with $\alpha=1 \times 10^{-3}$. Although different values of the $\alpha$ parameter can significantly influence the evolution of the disk, the timescale of gap opening is much shorter than the evolution timescale, so we did not include disk evolution in our code.
The disk extends from 0.5 to 30 au and is covered by a grid of 744 × 682 cells in the radial and azimuthal directions, respectively.
Damped boundary conditions are applied at both the inner and outer radial edges of the computational domain to prevent spurious wave reflections and ensure numerical stability. \\*
Dust particles are initially uniformly distributed within each ring, resulting in a radial density distribution that declines as 
 $1/r$. This distribution can be re-normalized to obtain any radial density profile for the dust. \\*
Initially, the planets are placed on circular orbits with a mass of 1 Earth mass, which increases to their final value over a ramp-up time of 100 yrs to avoid instabilities in the gas.

\section{The single planet case}
\begin{figure*} 
\centering
\includegraphics[width=\linewidth,height=0.22\textheight,keepaspectratio]{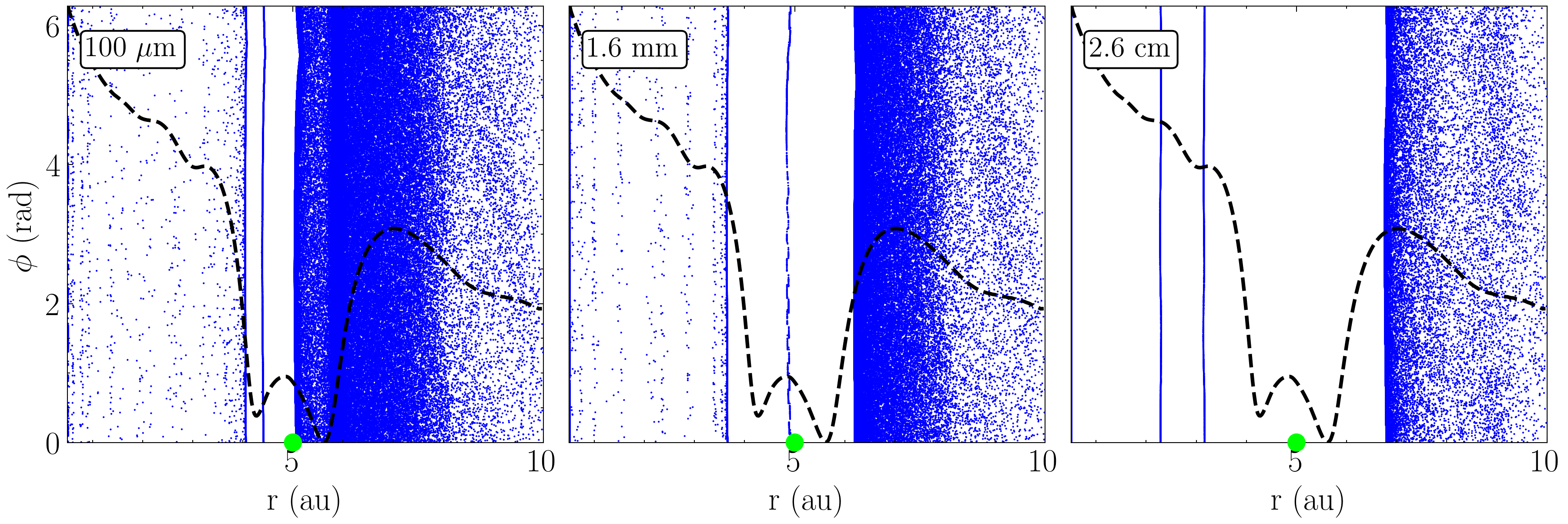}
\includegraphics[width=\linewidth,height=0.22\textheight,keepaspectratio]{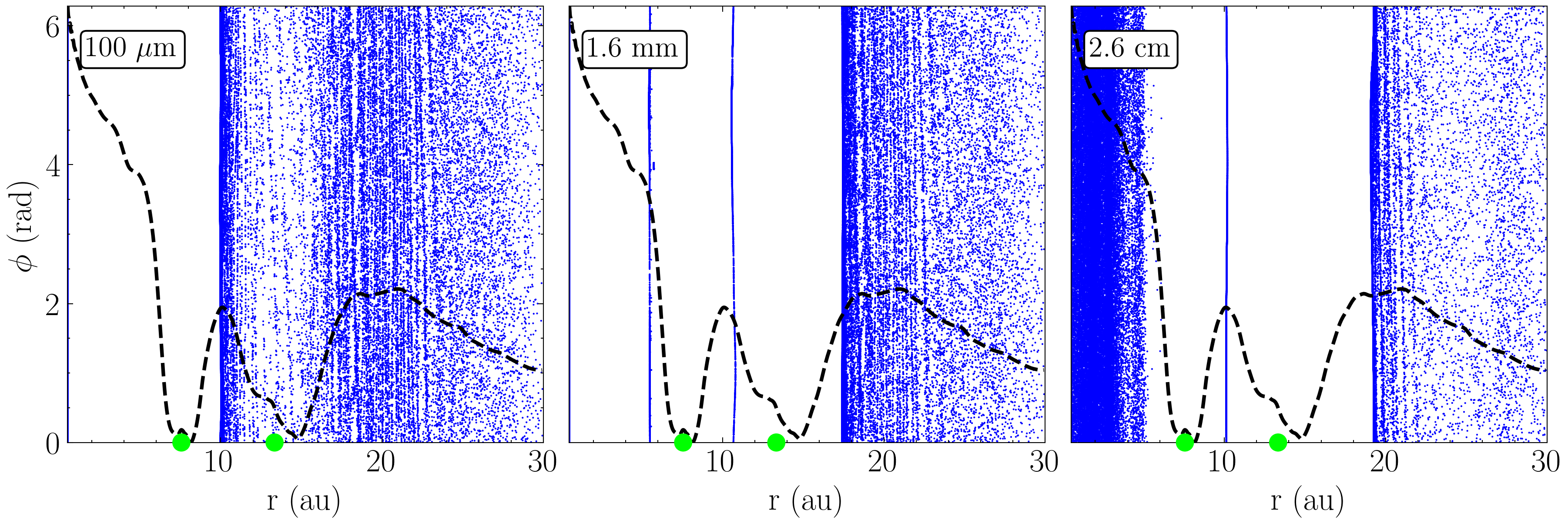}
\includegraphics[width=\linewidth,height=0.22\textheight,keepaspectratio]{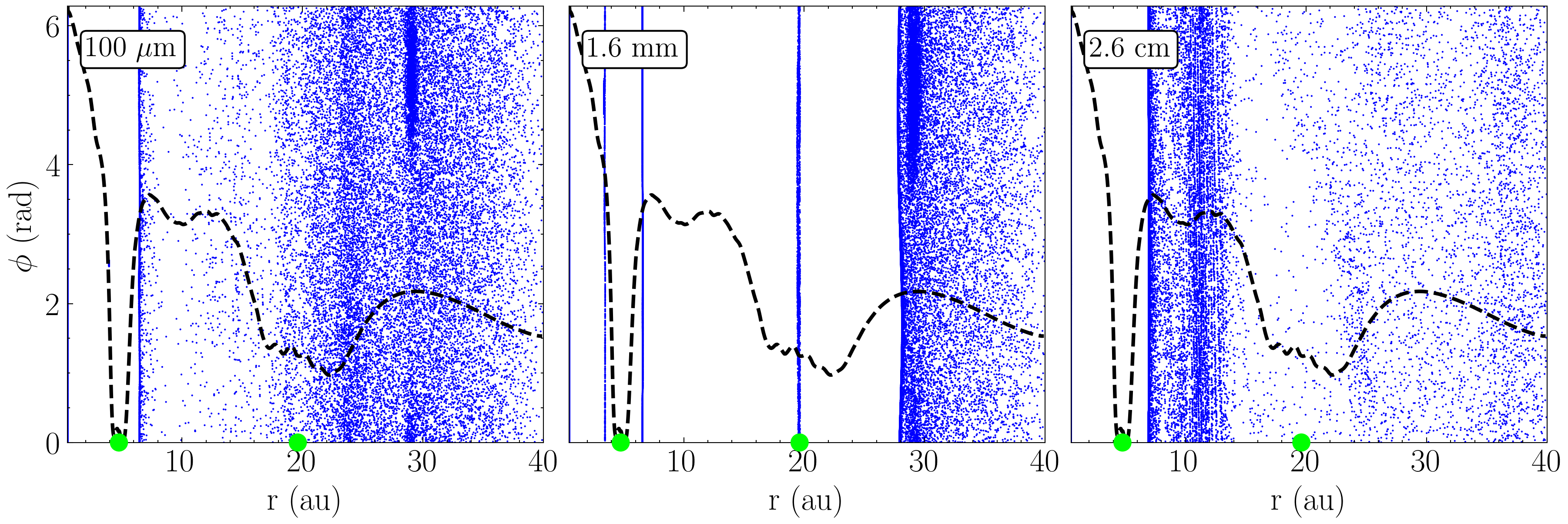}
\includegraphics[width=\linewidth,height=0.22\textheight,keepaspectratio]{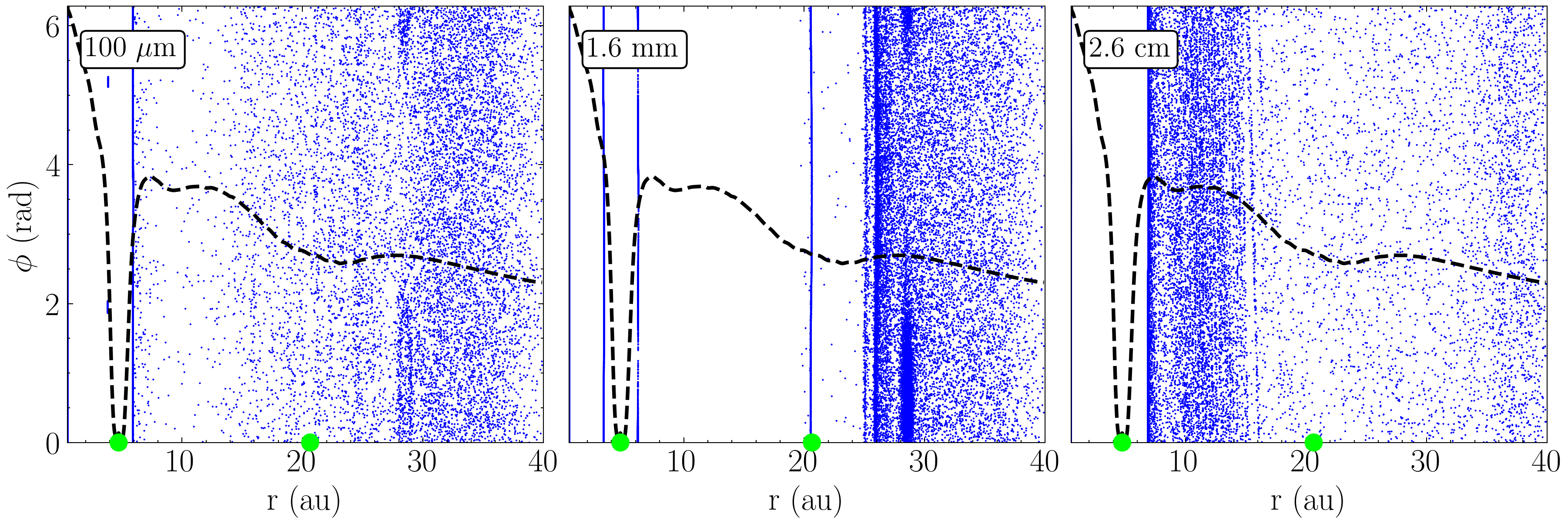}
\caption{
Dust distributions for three representative grain sizes. From left to right: 100 $\mu$m, 1.6 mm, and 2.6 cm particles. Rows correspond to different planetary configurations: single planet (top), two Jupiter-mass planets on close orbits (second), two Jupiter-mass planets on wide orbits (third), and a Jupiter–Saturn pair on wide orbits (bottom). The black dashed curve shows the radial gas surface density profile (normalized between 0 and 2$\pi$). Green filled circles mark the planet locations.
        }
\label{fig:1}
\end{figure*}

\begin{figure*} 
\centering
\begin{adjustbox}{width=1\linewidth}
\resizebox{\linewidth}{!}{%
\includegraphics[clip]{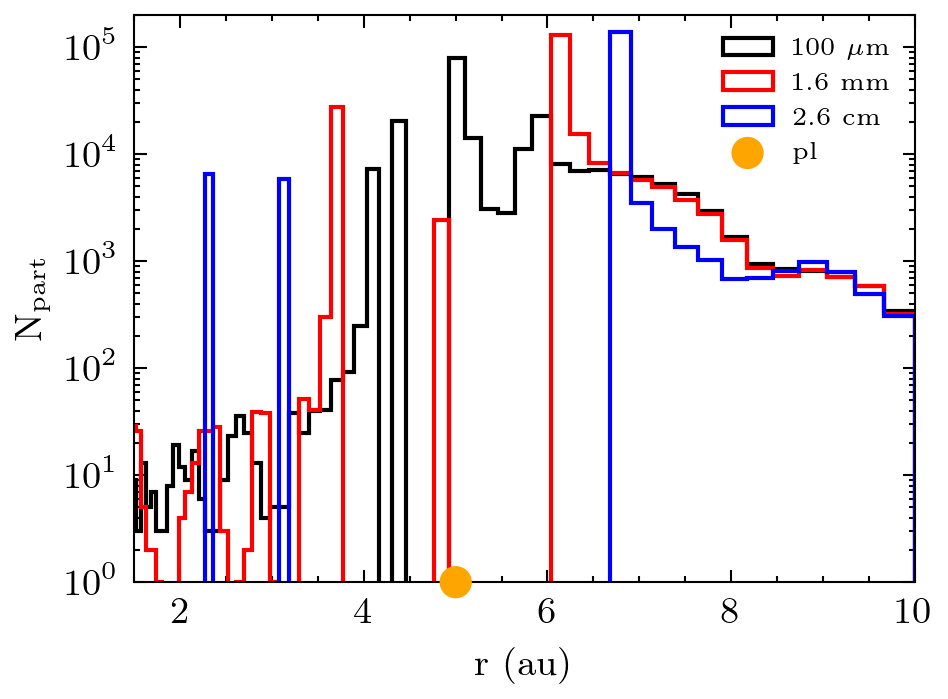}
\includegraphics[clip]{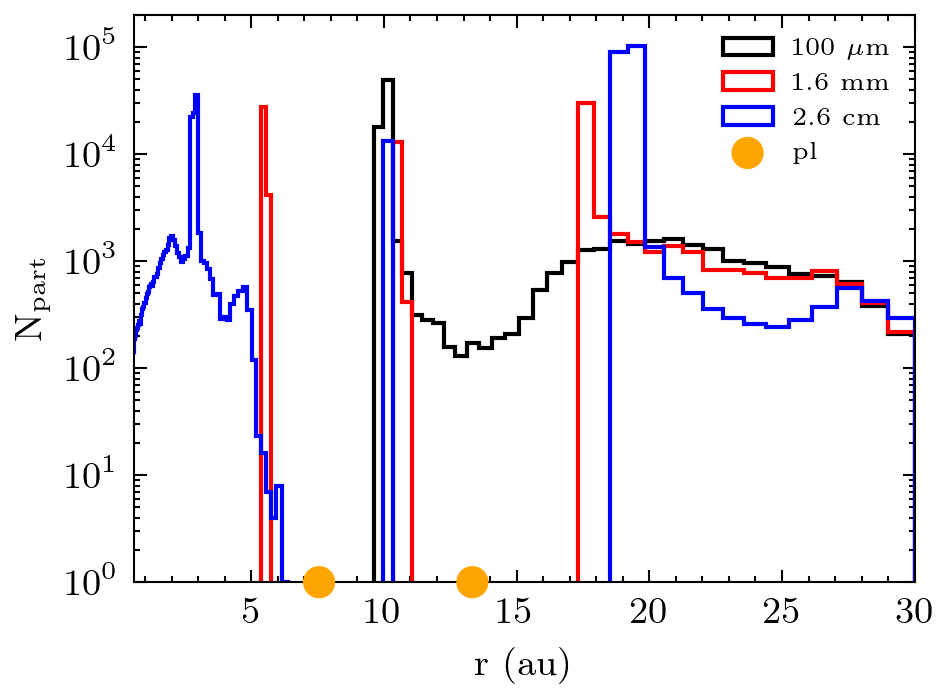}}
\end{adjustbox}
\begin{adjustbox}{width=1\linewidth}
\resizebox{\linewidth}{!}{%
\includegraphics[clip]{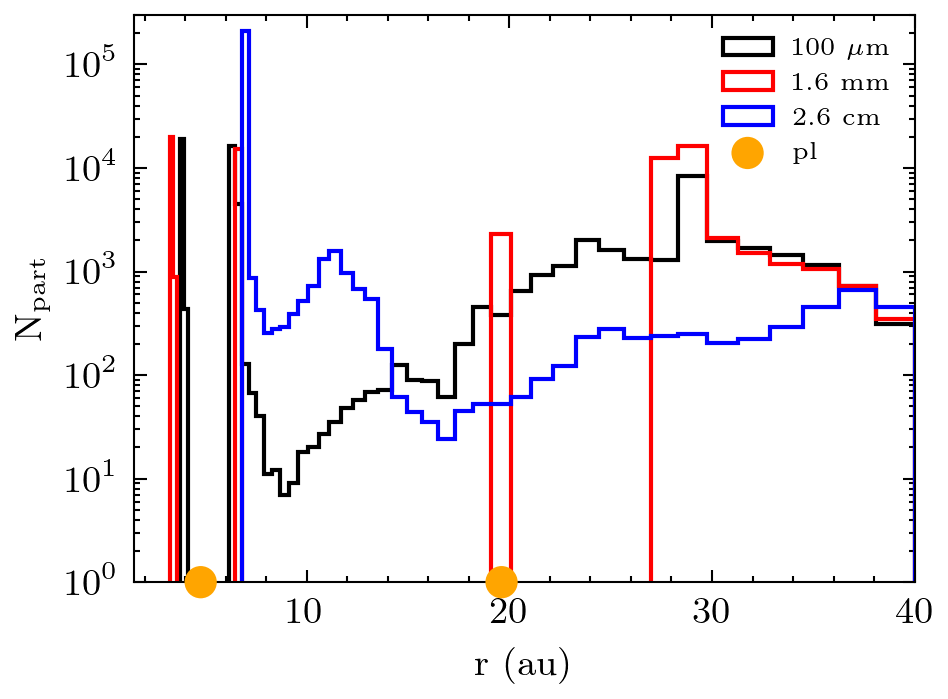}
\includegraphics[clip]{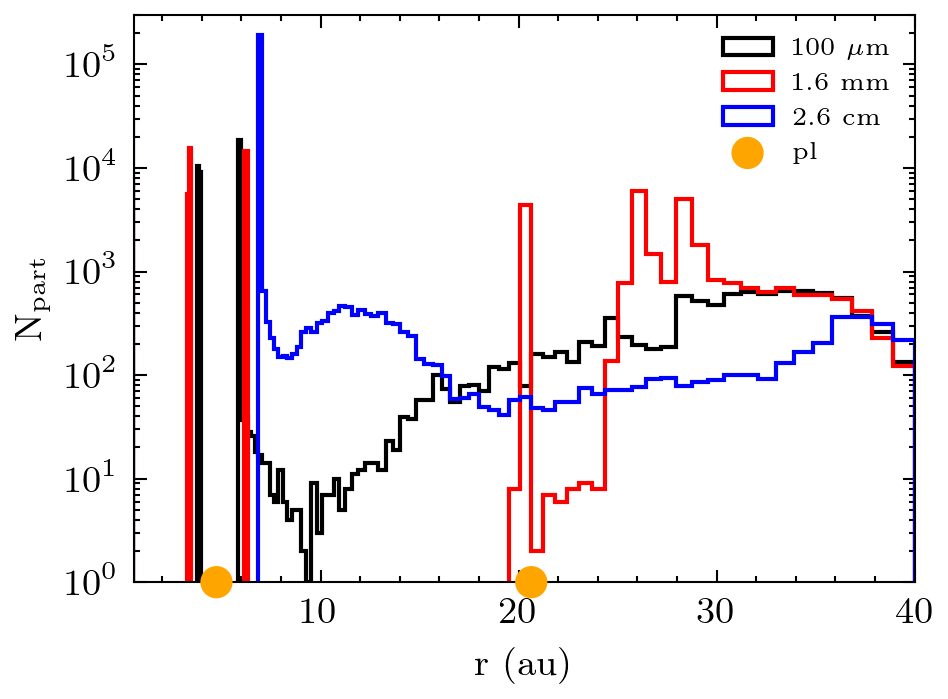}}
\end{adjustbox}
\caption{
Radial distribution of dust particle number density for each model, computed in radial bins of 0.25 au. Black, red, and blue lines correspond to 100 $\mu$m, 1.6 mm, and 2.6 cm particles, respectively. Orange filled circles mark the planet locations. Top left: single planet. Top right: two Jupiter-mass planets on close orbits. Bottom panels: wide-orbit configurations.
        }
\label{fig:hist}
\end{figure*}

As a preliminary test, we carried out a simulation with a single migrating giant planet of one Jupiter mass, initially located on a circular orbit at a distance of 5 au from the central star. \\*
The resulting dust distributions after 270 planet orbits are shown in {the top row of } \cifig{fig:1}. The planet-induced gap acts as a barrier for larger dust particles, as the mass of the planet exceeds the {so-called} pebble isolation mass. The latter is defined as the mass at which {the accretion of cm-sized dust grains ("pebbles")} stops, when the planetary core generates a pressure bump that traps drifting pebbles outside its orbit \citep{2014A&A...572A..35L, 2018A&A...612A..30B}
\begin{equation}
    M_{\rm{iso}} \approx 20 \left( \frac{a}{5\, \rm{au}} \right) ^{3/4} M_{\oplus},
\end{equation}
where the exact value also depends on the viscosity of the disk and the diffusion of dust particles, which varies with particle size \citep[see][]{2018A&A...612A..30B}.
However, smaller particles can filter through the outer dust trap. In particular, we observe small dust grains in the horseshoe region, but they could also represent particles that would have been accreted. In the histogram of \cifig{fig:hist} {(top left)}, it is also evident that the outer peak in dust density is more pronounced. This happens because in our simulations we have a constant flux of dust particles coming from the outside (i.e., from the outer disk) which are trapped at the outer edge of the dust gap, whereas particles in the inner region tend to drift towards the star and vanish from the simulation. Furthermore, the width of the dust gap depends on particle size, as the inward drift is halted at different equilibrium distances from the star. The wavy pattern of the gas distribution creates additional dust traps, particularly for pebbles, as shown in {the top row of } \cifig{fig:1}. {The formation of multiple dust traps by a single planet is consistent with previous results for a planet on a fixed circular orbit \citep{2018ApJ...866..110D} and for a migrating planet \citep{2020MNRAS.493.5892W}. In our case, the planet migrates slowly and multiple dust traps develop only for large particles with ${\rm St} \sim 1$, which are more efficiently concentrated at the pressure bump. Small particles can get closer to the planet, and their orbits are significantly excited (see the top left panel of \cifig{fig:ecc}).}
\section{Models with two Jupiter-mass planets}
If more than one planet is present in the system, the gravitational perturbations on the gaseous disk become complex and cannot be accurately modeled as a simple superposition of the individual effects from each planet. A similar situation applies to the dust distribution, which responds to the resulting complex gas dynamics and may accumulate in different gas-induced dust traps
or have its inward drift halted at different locations compared to the single-planet case. \\*
In the following, we consider three different scenarios. In the first, two Jupiter-sized planets orbit the star on relatively close orbits but remain far enough from the main first-order resonances. In the second scenario, the two planets are placed farther apart. In the third case, the mass of the outer planet is reduced to that of Saturn.
\subsection{Close-Orbit Configurations} \label{sec_jup_jup}
\begin{figure*}
\centering
\begin{adjustbox}{width=1\linewidth}
\resizebox{\linewidth}{!}{%
\includegraphics[clip]{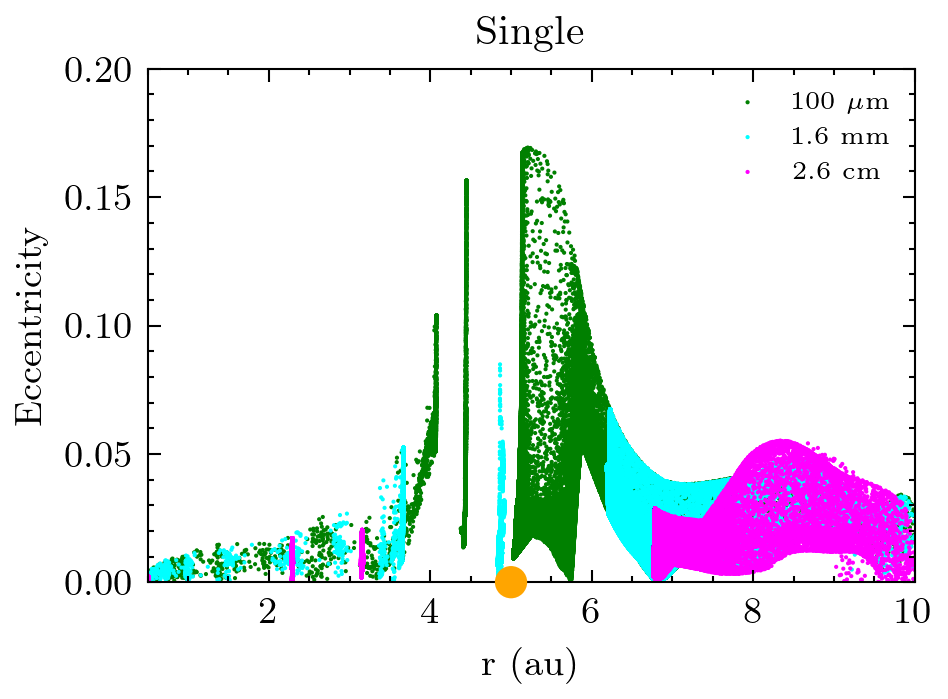}
\includegraphics[clip]{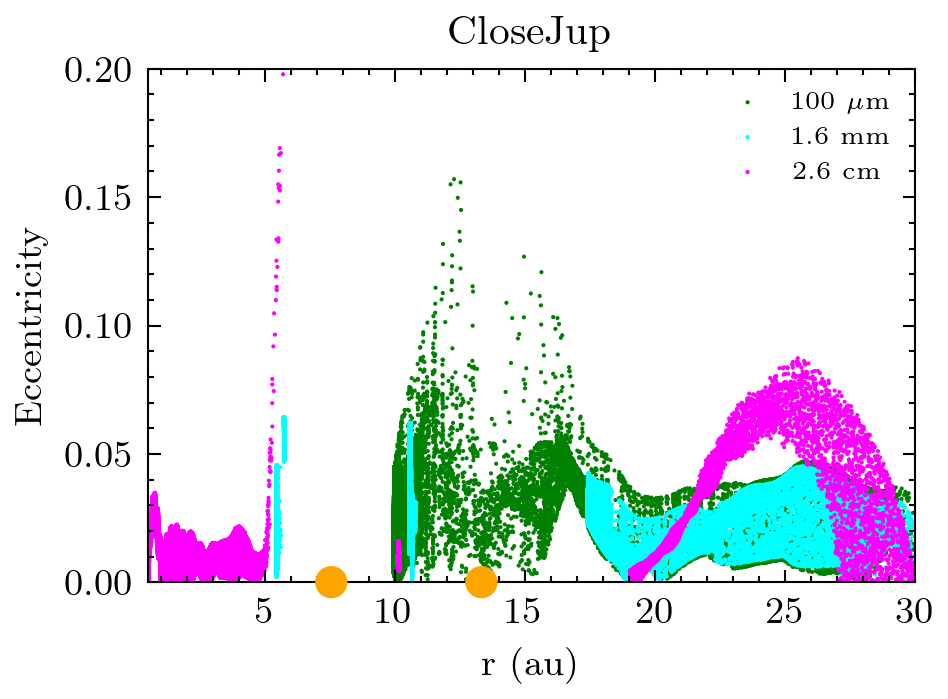}}
\end{adjustbox}
\begin{adjustbox}{width=1\linewidth}
\resizebox{\linewidth}{!}{%
\includegraphics[clip]{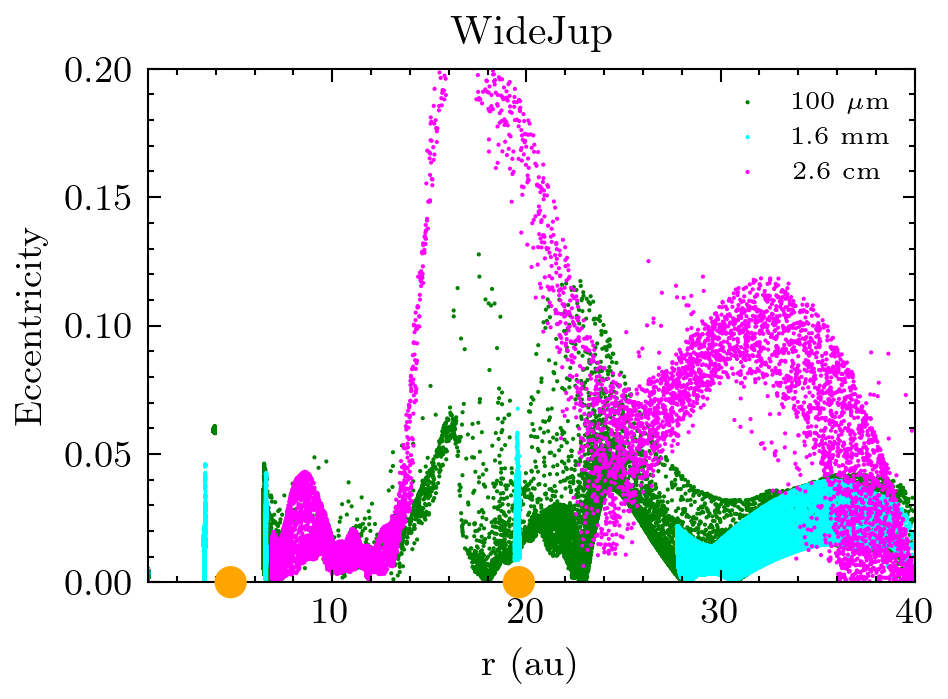}
\includegraphics[clip]{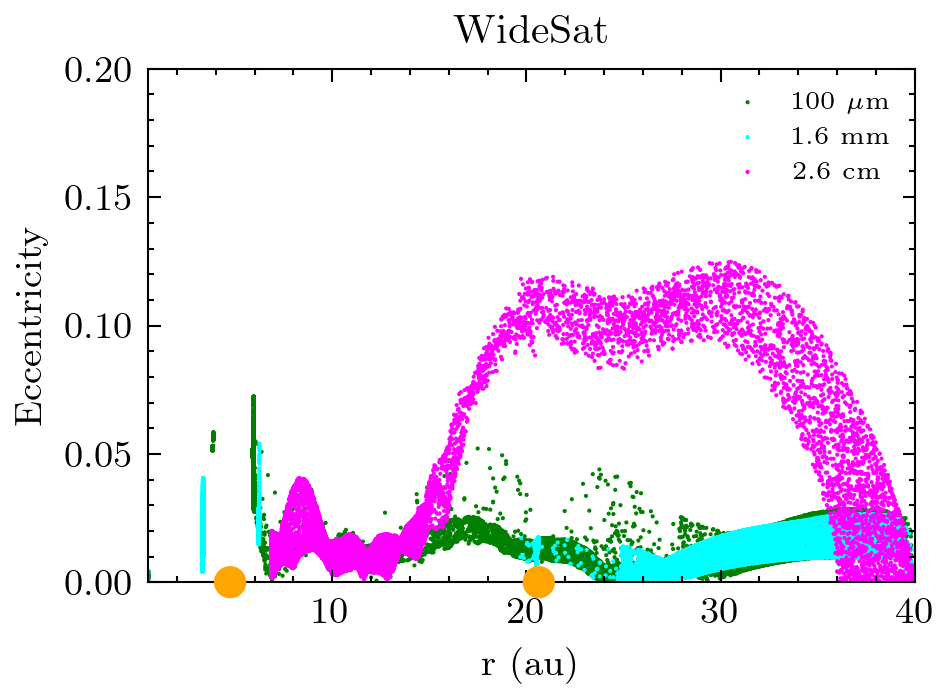}}
\end{adjustbox}
\caption{
Dust particle orbital eccentricity versus radial distance {for each model}. Green filled circles mark planet locations. Colored points denote grain sizes of 100 $\mu$m (green), {1.6 mm (cyan)}, and 2.6 cm (magenta).}
\label{fig:ecc}
\end{figure*}

We consider two Jupiter-sized planets: the first is initially placed on a circular orbit with a semi-major axis of 8 au, while the second orbits farther out, also on a circular orbit, with a semi-major axis of 15 au. \\*
{In the second row of} \cifig{fig:1}, we show the dust distribution for 100-$\mu$m, 1.6-mm, and 2.6-cm particles (pebbles) after 270 orbits of the outer planet.
Around the inner planet, the gap is well defined at all particle sizes, as also confirmed from the dust distribution histogram (\cifig{fig:hist}{, top right}). For the outer planet, the gap in the small particles is not as deep as expected, implying significant filtering of particles from the outer dust trap. 
The histogram clearly shows that the gap is continuously replenished by particles coming from outside. Additionally, notable structures appear in the regions beyond the outer dust trap because of spiral waves excited by the planet. As in the case of a single planet, the size of the gap depends on the size of the particles. Pebbles are trapped in the outer dust trap induced by the outer planet, forming multiple ring-like structures associated with local maxima and minima in the gas density. This ultimately produces a single wide gap that encompasses both planets. However, contrary to expectations, pebbles do not disappear inside the inner planet, thereby preventing the formation of a typical transition disk at this dust size. Most of the pebbles are concentrated near the water snowline located at 3.06 au, which could also play a role in slowing down these particles in the inner disk. The sudden change in the Stokes number of the particles creates a "traffic jam" mechanism where pebbles can accumulate. In contrast, a transition disk is formed for smaller dust sizes (left and middle panels {in the second row of} \cifig{fig:1}). These effects can likely be attributed to the small eccentricities acquired by the planets as a result of mutual secular perturbations. The inner planet reaches an average eccentricity of approximately 0.003, while the outer planet attains an average eccentricity of about 0.005 (in the single planet case it is 0.001). \\*
In addition, the {gas spirals} and {the} gravitational perturbations from both planets contribute to the excitation of the orbital eccentricities of the dust particles. {Far from the planets, smaller particles that are well coupled to the gas have their eccentricity primarily driven by the gas spiral; larger particles, instead, are dominated by the planet's gravitational perturbations; in the vicinity of the planets, the gravitational perturbations dominate across all particle sizes.} As shown in {the top right panel of} \cifig{fig:ecc}, the eccentricities of small 100 $\mu m$-sized particles are high within the gap opened by the outer planet. These particles are dragged into the gap by the gas spiral, and as a result they acquire significant eccentricity. For pebbles, the eccentricity is particularly high in the outer region of the disk, which may explain the observed drop in density near the outer dust trap. Because of their increased eccentricity, the particles migrate inward more rapidly as a result of enhanced drag with the gas. {This implies that the Stokes number of a particle of a given size decreases with increasing $\mathbf{v}_{\rm{rel}}$ (see eq. \ref{taus}), which should be taken into account when interpreting dust observations in terms of the Stokes number of the particles.}
\subsection{Distant Planetary Orbits} \label{sec_jup_jup_far}
\begin{figure}
    \centering
    \includegraphics[width=\linewidth]{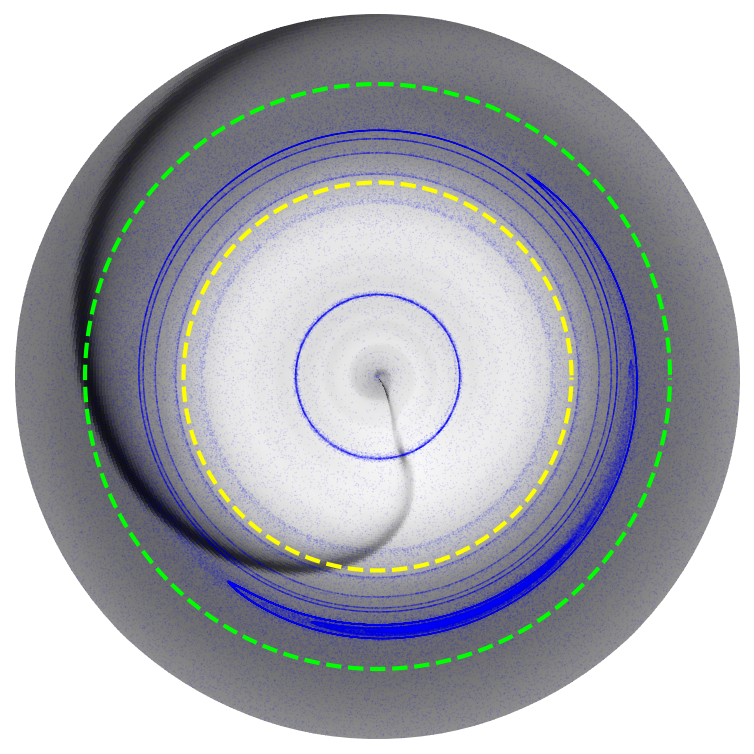}
    \caption{Gas and dust distributions for two widely separated Jupiter-mass planets. Blue dots represent {1.6-mm} dust particles, superimposed on the perturbed gas surface density $\Delta \Sigma / \Sigma _0$ (color scale). Green (yellow) dashed lines indicate the 2:1 (3:2) mean motion resonance with the outer planet. The plot is zoomed in the region between 15 to 35 AU to better visualize the dust overdensity near the outer gap edge.}
    \label{fig:snapshot}
\end{figure}

In this section, we consider a configuration with two Jupiter-sized planets placed farther apart than in the previous case. The inner planet is initially on a circular orbit with a semi-major axis of 5 au, whereas the outer planet orbits at a semi-major axis of 22 au. In this case, the disk extends to 40 au, and the grid is set to 796×546. The density and temperature profiles are identical to those in the previous model. The simulation ends after 270 orbits of the outer planet, as in the previous case. \\*
{The third row of }\cifig{fig:1} illustrates the dust distribution {for this model}. The 1.6 mm dust particles exhibit a behavior similar to that shown in the case of two closer planets: a single gap forms, which can be interpreted as due to a single massive planet. However, the pebble distribution is significantly different. In the case of close planets, a dense ring forms within 5 au, inside the orbit of the inner planet, whereas for two distant planets, the dense ring forms between the two, at distances between 6 and 14 au. As a consequence, a completely different interpretation can be derived simply by examining the pebble distribution in terms of the size and location of the planets responsible for the formation of the rings. \\*
The distribution of smaller particles also appears to be significantly different. With close planets, there is an inner gap and a mild gap around the outer planet. In contrast, when the planets are more widely separated, a gap appears between the two planets. This is confirmed by inspecting the histograms in \cifig{fig:hist} {(top right and bottom left panels)}. \\*
In addition, in \cifig{fig:1} we identify a pronounced overdensity of small and intermediate-size dust grains at a radial distance of approximately 30 AU, which is clearly non-axisymmetric. To investigate the origin of this feature, we present in \cifig{fig:snapshot} the spatial distribution of dust particles overlaid on the perturbed gas surface density, $\delta \Sigma / \Sigma_0$. We find no evident spatial correlation between the dust overdensity and the spiral wake launched by the outer planet. Moreover, we do not detect the presence of a vortex or localized pressure maximum that could be associated with the Rossby wave instability. The dust overdensity is radially narrow, suggesting a possible connection with mean-motion resonances (MMRs) with the outer planet. However, we find that its location does not coincide with a specific resonance but instead lies between the nominal 2:1 and 3:2 MMRs. Instead, the radial location of the dust overdensity is related to the pressure bump generated by the outer planet, which traps the small and intermediate-sized dust particles, whereas pebbles are able to filter through the gap and are trapped in the pressure bump generated by the inner planet. {This azimuthal asymmetry is present only in the dust at this stage, but we identify in earlier snapshots a perturbation in the gas vortensity consistent with transient vortex formation. As the planets migrate inward, the gas feature dissipates, while dust remains trapped for longer timescales. } Interestingly, this asymmetric structure is absent both in the single planet case and the two Jupiter-mass planets in close orbits of the previous Section. This may be due to dust diffusion, which tends to smooth out dust perturbations and acts on a shorter timescale in these cases. \\*
Finally, the arch shape observed in \cifig{fig:1} for the 2.56 cm dust particles is due to the high eccentricity excited by the secular perturbations of the two planets, combined with the gas perturbations. {In the bottom left panel of \cifig{fig:ecc}, we show the eccentricity distribution of the same particles in this model. Close to the outer planet, eccentricities as high as $e \sim 0.2$ are reached, leading to complex dynamics in its vicinity and to a non-trivial radial distribution. Even in this case, the elevated eccentricities make a classification of particles based solely on their Stokes number problematic, since the dust–gas drag depends on the instantaneous relative velocity along the orbit.\\*
For particles on eccentric orbits, $\mathbf{v}_{\rm{rel}}$ is systematically larger than in the circular case, owing to both radial and azimuthal velocity differences between the particle and the gas. As a consequence, the effective coupling strength averaged over an orbital period differs from that inferred under the assumption of circular motion.\\*
This has two important implications. First, when simulation results are reported only in terms of a Stokes number, without specifying the orbital elements (in particular the semi-major axis and eccentricity), the corresponding physical grain size cannot be uniquely determined. In eccentric configurations, the mapping between ${\rm St}$ and grain size depends not only on the local gas properties but also on the orbital parameters. Second, observations generally do not constrain particle eccentricities directly. Therefore, adopting a Stokes number derived under the assumption of circular motion may bias estimates of the dust–gas coupling and, consequently, the planetary mass required to reproduce the observed disk structure.}
\section{Models with two planets of different mass} \label{sec_jup_sat}
Another interesting case involves two planets of different mass, modeled after Jupiter and Saturn. As in the previous setup, the inner planet is initially placed on a circular orbit with a semi-major axis of 5 au, while the outer planet starts with a semi-major axis of 22 au. The disk parameters remain the same as in the earlier model, and the simulation is terminated after 270 orbital periods of the outer planet. \\*
By comparing {the third and fourth rows of } \cifig{fig:1}, we can observe morphological differences in the dust distribution compared to the case with two Jupiter-mass planets. For small grains of 100 $\mu$m, the distribution is broadly similar, with the exception of a gap located closer to 15 au rather than 20 au. {In the distribution of 2.6-cm pebbles, the wavy pattern appears less pronounced in the case of a Saturn-mass planet, as it produces weaker gas waves}. For intermediate-sized particles (1.6 mm), the dust distribution follows that of {the previous model}, but the inner edge of the large inner hole extends to 25 au, whereas in the case with two massive planets it is located at 28 au. This clearly poses a challenge when trying to interpret the size of a large inner hole in the dust distribution of a disk in terms of planet mass. Such a feature could be produced either by two nearby massive planets or by two planets where the outer one is less massive and located farther away. Then, for pebbles, the overall distributions remain similar, although beyond 15 au the dust density is lower than that with massive planets. \\*
In addition, we note the presence of an asymmetric structure in the dust distribution near the edge of the outer gap, in particular for millimeter-sized grains. Similarly to the previous case, most of the grains are trapped in the pressure bump generated by the outer planet, whereas pebbles are able to escape because of their weak coupling with the gas. {As mentioned before,} the asymmetry is {likely} due to a vortex that has not been smoothed out by particle diffusion; moreover, the overdensity is less prominent in the small grains compared to the previous simulation, {likely} because of the lower mass of the outer planet. These features are confirmed by \cifig{fig:hist} {(bottom right panel)}, where the density of the dust particles shows that intermediate-size grains are mostly stopped at the outer edge of the outer planet's orbit creating a large cavity, whereas both smaller and bigger grains are able to filter through the gap: in particular, 2.6 cm pebbles are mostly concentrated in between the two planetary orbits. \\* 
By comparing {the dust distributions for each model in \cifig{fig:1}}, together with the corresponding histograms in \cifig{fig:hist}, it appears difficult to identify a common behavior. For most particle sizes, a transition disk forms which in some cases may extend slightly beyond the orbit of the inner planet, while in others it may extend outside the orbit of the outer planet. This leads to an imbalance in the dust size distribution, which is not homogeneous but instead depends on the radial distance from the star, in a manner consistent with our findings for sub-thermal mass planets in \cite{2025A&A...703A.270R}. This complicates the interpretation of real disk observations, where a broad range of dust sizes with different optical depths contributes to the emission, potentially blending or masking the dynamical signatures produced by embedded planets. 
\section{Synthetic images}
{To evaluate the observational impact of our results, we generated ALMA-like synthetic continuum images for each simulation. We show in \cifig{fig:raw_brightness} the raw emission maps (before beam convolution) for the different multi-planet systems simulated in Band 7 (0.85 mm), Band 6 (1.3 mm) and Band 3 (3 mm).} To have a more realistic picture, we created mock observations by convolving the raw emission maps with realistic ALMA beam sizes: configurations C43-8, C43-9, and C43-10 were used for Bands 7, 6, and 3, yielding beams of 0.028", 0.024", and 0.041", respectively.
\begin{figure*} 
    \centering
    \includegraphics[width=\linewidth]{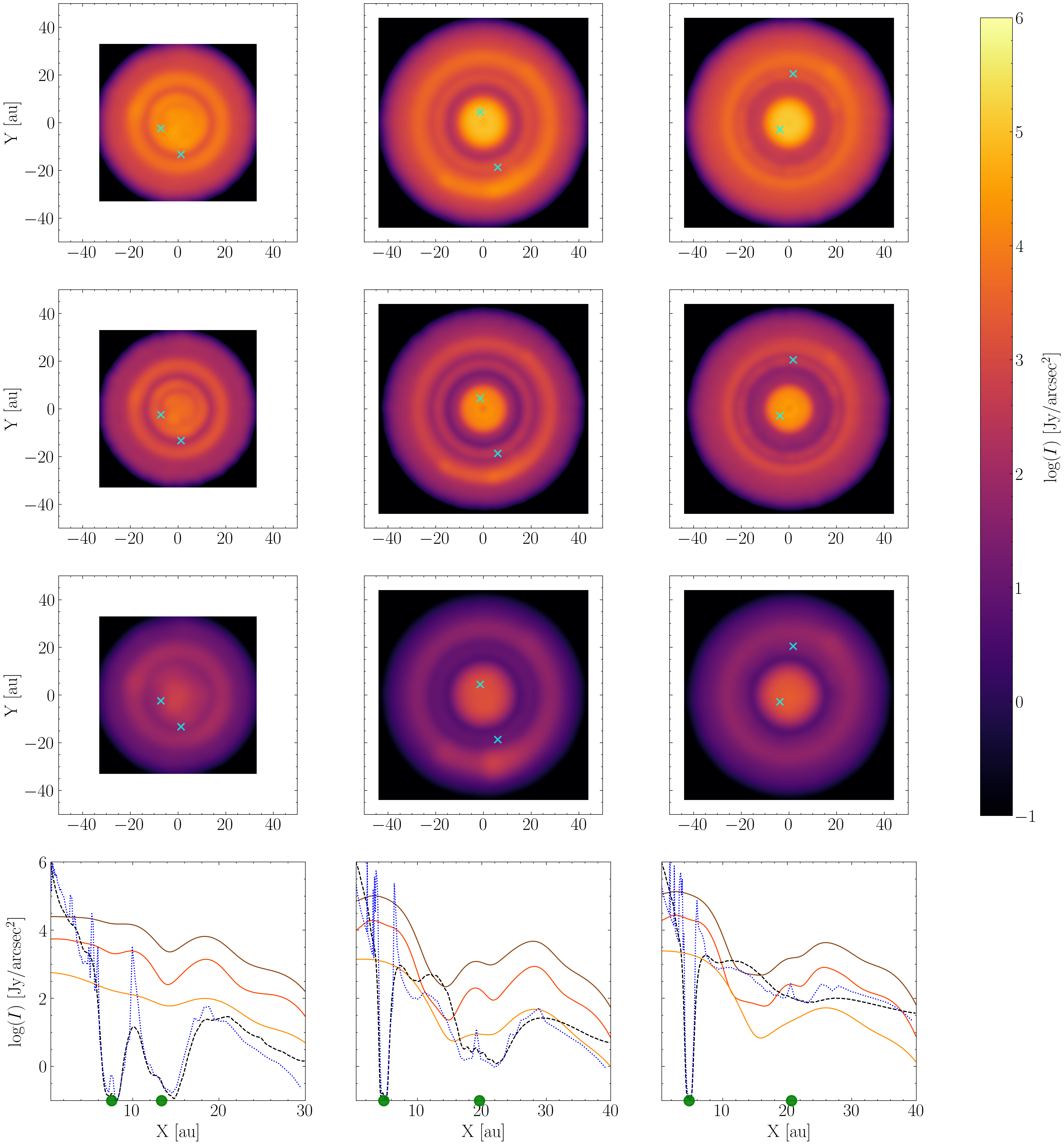}
    \caption{Dust continuum emission convolved with a realistic ALMA beam in Band 7 (first row), Band 6 (second row) and Band 3 (third row). Columns correspond to the models CloseJup (left), WideJup (middle), and WideSat (right). Bottom panel: radial intensity profiles in Bands 7 (brown), 6 (orange), and 3 (yellow), together with dust (blue dashed) and gas (black dashed) radial profiles. Cyan crosses and green filled circles mark the planet locations.}
\label{radmc}
\end{figure*}
\cifig{radmc} shows the resulting intensity maps (Jy/arcsec$^2$) for each system, along with radial profiles of dust and gas (dashed lines). The intensity is generally higher at shorter wavelengths, indicating that small grains dominate the continuum emission. In most cases, the inner planet is barely visible due to its proximity to the star and the narrowness of its dust gap, which is unresolved by the beam. In the close-orbit system (left column), the inner gap is detectable only in Band 6, while the outer gap is visible in all bands. In wide-orbit systems, only one gap appears in between the two planets. This is because continuum emission arises from all dust sizes, each with a markedly different spatial distribution (see histograms in \cifig{fig:hist}), so that the combined signal rarely produces two clearly separated gaps. This shows the challenge of interpreting the dust distribution shaped by more than one planet. In particular, inspection of the images by themselves provides very limited constraints on the planetary masses and semi-major axes, and even the number of planets present in the disk cannot be robustly determined. \\*
In addition, asymmetric structures are observed in some cases, particularly for two distant Jupiter-mass planets (center column, near 6 o'clock), outside the orbit of the outer planet. These correspond to a local dust overdensity that originates from hydrodynamical simulations and is smoothed out by the beam. This feature is most prominent in Band 6, showing that intermediate-sized grains are more concentrated in the clump, whereas smaller and larger grains are more diffused, producing more extended emission in Bands 7 and 3 (as seen in \cifig{fig:1}). Such structures resemble crescent-shaped asymmetries seen in disks such as MWC 758 \citep{2019MNRAS.486..304B}. {However, at this stage, we do not detect a strong gas asymmetry that would indicate an active Rossby wave instability; in our simulations, the dust clump is consistent with the remnant of an earlier perturbation in the gas vortensity, near the outer pressure bump. As the planets migrate inward, the original gas feature dissipates, while the dust remains trapped on longer timescales; this dust concentration develops after $\sim 5$ kyr in both wide-orbit configurations, whereas it does not form in the case of close planets.}
\subsection{Planetary mass estimation with DBNets2.0}
We claimed that a multi-planet system cannot be interpreted as a mere superposition of the effects of the single planets, so that estimating the planetary mass assuming that it acts alone may lead to inaccurate conclusions. To verify this, we fed our mock observations to DBNets2.0 \citep{2025A&A...700A.190R}\footnote{We used the python tool available at \url{https://github.com/dust-busters/DBNets/tree/dbnets2.0.0}}: a deep learning tool, based on convolutional neural networks (CNNs), that analyzes dust substructures observed in the continuum emission to infer the mass of putative embedded planets, together with disk properties such as $\alpha$-viscosity, the disk aspect ratio at the planet location $h_p$ and the dust Stokes number $St$ . This tool is built to find single planets, although one could apply it multiple times to find a planet for each observed gap. Because we observe only one gap in each system, we applied the tool to search for a planet at the gap location (as an observer would do with a real image), and we repeated this procedure only for Band 6 and 7, as they are compatible with the training set \citep{2025A&A...700A.190R}. \\*
The results are shown in Table \ref{tab:dbnets}: for each measurement, we also show the confidence score (CS), a metric that helps to assess the reliability of the inferred posteriors, particularly when applying the tool to out of distribution data. We find CS > 0.6 for all configurations, so that all estimates can be considered acceptable (\cite{2025A&A...700A.190R}). For the system with two Jupiter-mass planets in a close orbit, a narrow gap is observed near the location of the outer planet, so that its mass is in the range measured by DBNets, although slightly underestimated. In the wide-orbit configurations, a wide gap is observed between the two planets: as a consequence, the tool finds a higher planetary mass. This shows how the interpretation of observed gaps requires careful consideration: even if a dust gap seems to be produced by a single planet, it can be the outcome of a multi-planet system embedded in the disk.
\begin{table}
\caption{Results of planetary mass estimation with DBNets2.0.}
\label{tab:dbnets}
\centering
\begin{tabular}{ c  c  c  c  c  }
\hline\hline
{Model} & {Band} & {r$_{\rm{gap}}$ (au)} & {M$_{\rm{DBN}}$ (M$_{j}$)} & {CS}  \\
\hline
CloseJup & 7  & 13  & 0.88$^{+0.41} _{-0.30} $ & 0.75\\
CloseJup & 6  & 12.79  & 0.86$^{+0.36} _{-0.29} $ & 0.75\\
WideJup & 7  & 13.11  & 2.83$^{+0.66} _{-0.52} $  & 0.84 \\
WideJup & 6  & 13.5  & 3.35$^{+0.98} _{-0.68} $  & 0.87 \\
WideSat  & 7  & 14.28  & 2.93$^{+0.52} _{-0.49} $ & 0.84 \\
WideSat  & 6  & 14.95  & 2.62$^{+0.69} _{-0.57} $ & 0.86\\[2pt]
\hline
\end{tabular}
\end{table}

\section{Dust impact velocities}
\begin{figure}
\centering\resizebox{\linewidth}{!}{\includegraphics[clip]{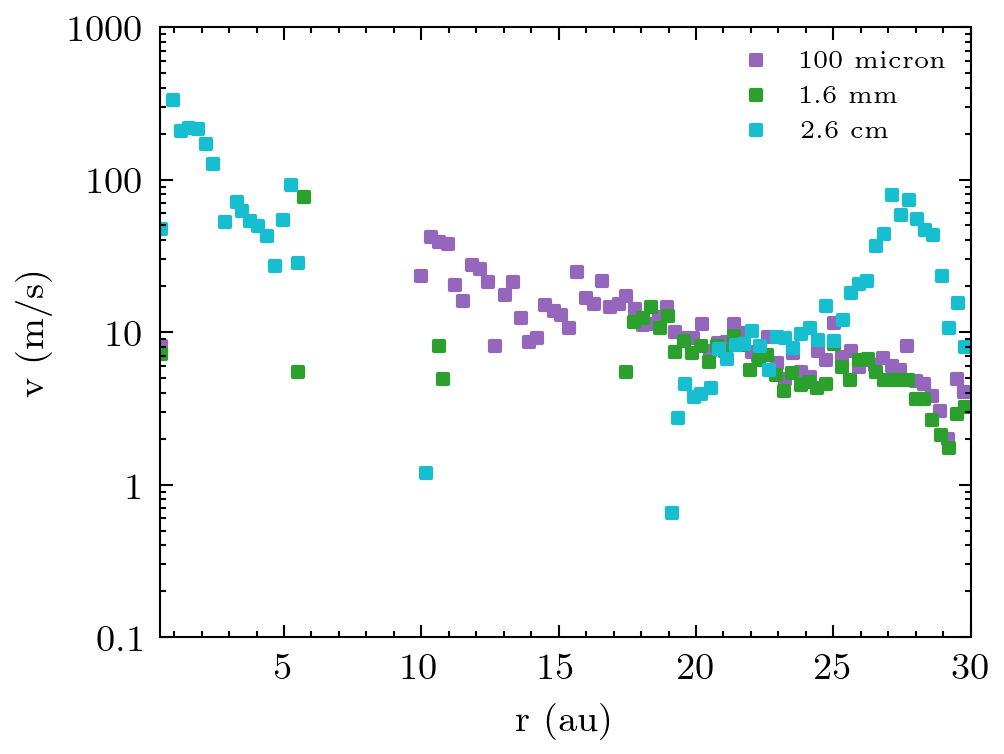}}
\centering\resizebox{\linewidth}{!}{\includegraphics[clip]{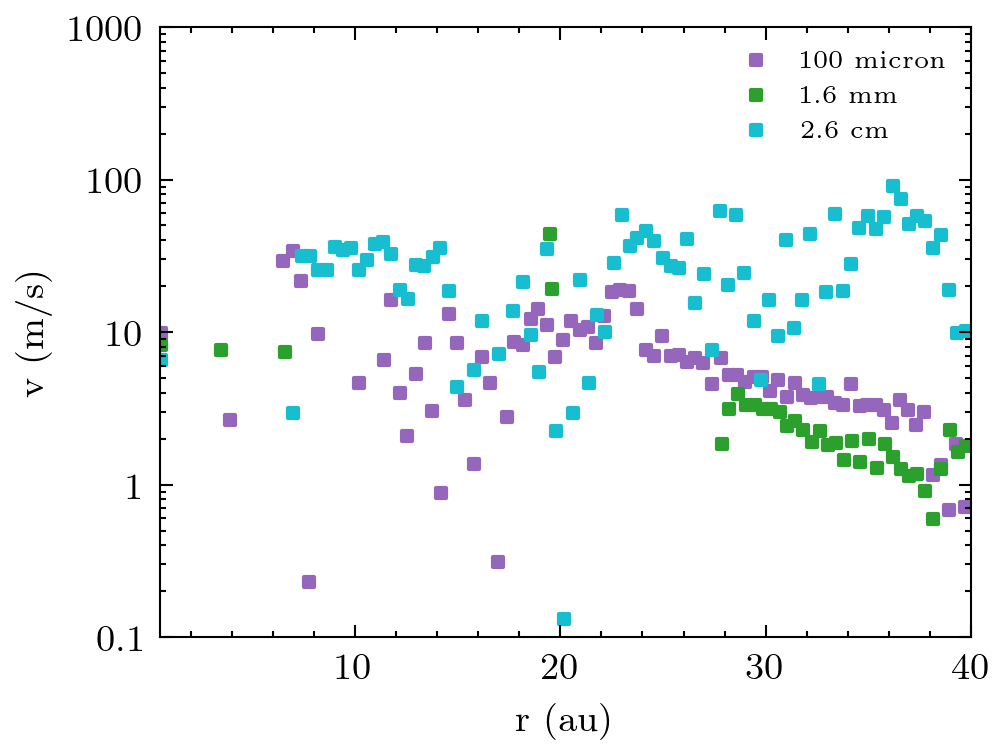}}
\centering\resizebox{\linewidth}{!}{\includegraphics[clip]{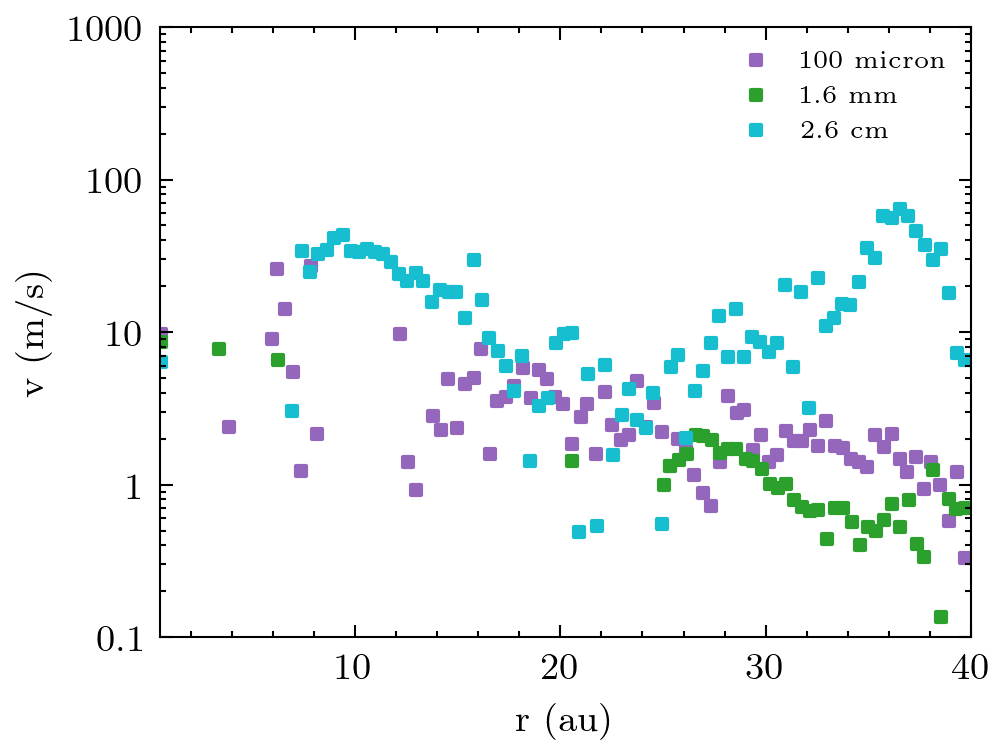}}
\caption{
Mean impact velocities between dust grains of the same size as a function of radial distance. Top: two Jupiter-mass planets on close orbits. Middle: two widely separated Jupiter-mass planets. Bottom: Jupiter–Saturn pair.
        }
\label{fig:vel}
\end{figure}

After the possible rapid formation of giant planets, dust accumulation and growth can continue within the disk, leading to the formation of additional planetesimals and planets. An important parameter in this scenario is the relative impact velocity between dust particles. In our simulations, we compute these relative velocities from the simulated particles by dividing the entire 2D disk into sectors defined by bins in radial distance and azimuthal angle. Within each sector, the relative velocities are calculated after subtracting the local Keplerian velocity, thereby removing the contribution from Keplerian shear. \\*
Since we have a large number of particles, we can keep the sectors sufficiently small to obtain a good approximation of the relative impact velocity due to the eccentricity of the orbits. \\*
Because the simulations are 2D, we neglect the contribution from the out-of-plane velocity component produced by the inclination excited by the gravitational perturbations of the planets. However, we can estimate its effect using the equation of \citet{lissauer1993}, which gives the relative velocity between planetesimals moving on orbits with average eccentricity $e$ and inclination $i$: 
\begin{equation}
v_{\rm rel} \approx v_K \sqrt{\frac{5}{8} e^2 + \frac{1}{2} i^2} ,   
\end{equation}
where $v_k$ is the local Keplerian velocity. The effect of a dispersion on the inclination is approximately equivalent to that of eccentricity, and it would increase the relative velocity by a factor of about $\sqrt{2}$. \\*  
We show in \cifig{fig:vel} the mean impact velocities computed between dust grains of the same size as a function of the radial distance from the star for each of the simulated multi-planet systems. Because of the gravitational perturbations of the planets and the resulting eccentric orbits, the relative velocities are on the order of a few m/s for 100-micron and 1.6-mm particles, and tens of m/s for pebbles. Whether such velocities allow dust growth depends critically on the fragmentation threshold velocity, $v_{\rm frag}$, which remains uncertain. Typical values of $v_{\rm frag}$ are often assumed to be on the order of a few m/s for silicate grains and somewhat higher for icy particles, but only if the local disk temperature is sufficiently high (see the review by \citealt{2024ARA&A..62..157B} and references therein). Recent work suggests that values up to $\sim$30 m/s may be possible depending on grain porosity and monomer properties \citep{2025A&A...703A.180P}.\\*
For fragmentation thresholds below this upper limit, the collision velocities found in our simulations would likely hinder further growth of dust into larger bodies, even within dust traps at the edges of planet-induced gaps where the dust density is enhanced. In this case, dust pile-ups would not necessarily translate into efficient growth, as collisions may be dominated by fragmentation rather than sticking, potentially limiting the formation of second-generation planetesimals. An exception may occur in regions where giant-planet perturbations generate local turbulence at gap edges. There, strong pressure and vortensity gradients may trigger vortex formation, which could concentrate solids and enhance collisional growth. \\*
From a theoretical standpoint, the dust content of circumstellar disks is expected to steadily decrease with age as micrometer-sized grains grow into larger aggregates, planetesimals, and eventually planets. However, an analysis of ALMA observations by \cite{testi2022} suggests that disks in star-forming regions with typical stellar ages of $\lesssim$ 1 Myr exhibit a similar or even lower median dust content compared to older disks with ages of 2–3 Myr. According to \cite{turrini2019} and \cite{barnabo2022}, this behavior may be attributed to an enhanced collisional evolution of planetesimals driven by planetary perturbations, which can create conditions for the collisional rejuvenation of the dust population, partially halting or even reversing the expected decline trend and significantly contributing to the observed dust-to-gas ratio.\\*
When operating in conjunction with planetesimal fragmentation, the collisional grinding of dust induced by the enhanced dust eccentricities discussed here may further amplify the observed resurgence of millimeter and sub-millimeter dust densities reported by \cite{testi2022}.
\section{Conclusions}
This study {highlights the challenges} that may arise when interpreting the dust distribution in a circumstellar disk shaped by more than one (super-thermal mass) planet.
\begin{itemize}
    \item Our numerical simulations show that, within the parameter range explored, the formation of clean, well-separated gaps almost never occurs across all dust sizes when two perturbing planets are embedded in the disk. This outcome arises from the complex wavy structures induced in the gas density, which generate multiple dust traps, {and from gravitational perturbations associated with the planets' eccentricity evolution, which excite the orbits of the dust particles.} Together, these two mechanisms strongly affect dust dynamics, resulting in markedly different size-dependent distributions (see \cifig{fig:1}). {These findings build upon previous studies of single-planet systems \citep{2018ApJ...866..110D, 2020MNRAS.493.5892W} by showing that the presence of a second perturber can substantially modify the orbital evolution of dust grains and produce disk substructures that are more complex and potentially ambiguous to interpret, even when the planets are widely separated.};
    \item The dust eccentricity enhances the headwind experienced by the particles. As a result, the drift rate is no longer determined solely by grain size and gas density, but is strongly dependent on the grains' eccentricity (see \cifig{fig:ecc}). {The latter significantly changes the relative velocity between the gas and the dust particles of a given size, reducing their Stokes number (see eq. \ref{taus}), which is crucial for the interpretation of observed dust distributions in terms of St.};
    \item From the results of our simulations, we generate realistic mock observations with RADMC-3D \citep{2012ascl.soft02015D}, including a convolution with an ALMA-like beam. This mock data reveals the intrinsic complexity of disks containing multiple planets and illustrates the challenges associated with inferring the physical and dynamical properties of embedded planets from observations alone (see \cifig{radmc}). To verify this, we tackle the inverse problem by using our mock data as input to estimate the planetary mass with DBNets2.0 \citep{2025A&A...700A.190R}: because there is only one visible gap in our images, an observer interprets it as due to a single planet located in the center of the gap and misestimates its mass (see Table \ref{tab:dbnets}). This shows how multi-planet systems complicate the interpretation of observed dust substructures: {a similar result on two planets showing a single observable gap has been found by \cite{2015A&A...573A...9P} using a 1D model for dust evolution and assuming a fixed circular orbit for the planets. We extended this result to migrating planets with different masses and orbital configurations, providing a more detailed view of the 2D dust distribution and kinematics for different grain sizes;}
    \item Dust particles experience enhanced relative velocities as a direct consequence of their eccentricities. This leads to collision velocities that can become comparable to, or even exceed, the critical threshold above which fragmentation dominates over grain growth and coagulation into larger aggregates (see \cifig{fig:vel}). As a result, the dust population undergoes progressive collisional grinding, effectively reversing the grain growth process and giving rise to an increasing abundance of smaller dust particles. This effect is also relevant in the vicinity of gap edges, where the dust surface density is highest and collisional rates are enhanced.
    This mechanism may contribute to sustained and continuous replenishment of small dust grains within the disk.
\end{itemize}

\begin{acknowledgements}
{We thank the anonymous referee for their comments and suggestions, which helped to improve the clarity and quality of this work.}
This publication was produced while attending the PhD program in Astronomy at the University of Padova, Cycle XXXIX, with the support of a scholarship co-financed by the Ministerial Decree no. 118 of 2nd March 2023,760 based on the NRRP funded by the European Union - NextGenerationEU - Mission 4 Component 1 – CUP C96E23000340001. 
\end{acknowledgements}
\bibliographystyle{aa.bst}

\bibliography{local}

@ARTICLE{2018ApJ...869L..41A,
       author = {{Andrews}, Sean M. and {Huang}, Jane and {P{\'e}rez}, Laura M. and {Isella}, Andrea and {Dullemond}, Cornelis P. and {Kurtovic}, Nicol{\'a}s T. and {Guzm{\'a}n}, Viviana V. and {Carpenter}, John M. and {Wilner}, David J. and {Zhang}, Shangjia and {Zhu}, Zhaohuan and {Birnstiel}, Tilman and {Bai}, Xue-Ning and {Benisty}, Myriam and {Hughes}, A. Meredith and {{\"O}berg}, Karin I. and {Ricci}, Luca},
        title = "{The Disk Substructures at High Angular Resolution Project (DSHARP). I. Motivation, Sample, Calibration, and Overview}",
      journal = {\apjl},
         year = 2018,
        month = dec,
       volume = {869},
       number = {2},
          eid = {L41},
        pages = {L41},
          doi = {10.3847/2041-8213/aaf741},
archivePrefix = {arXiv},
       eprint = {1812.04040},
 primaryClass = {astro-ph.SR},
       adsurl = {https://ui.adsabs.harvard.edu/abs/2018ApJ...869L..41A}
}

@ARTICLE{2018ApJ...866..110D,
       author = {{Dong}, Ruobing and {Li}, Shengtai and {Chiang}, Eugene and {Li}, Hui},
        title = "{Multiple Disk Gaps and Rings Generated by a Single Super-Earth. II. Spacings, Depths, and Number of Gaps, with Application to Real Systems}",
      journal = {\apj},
         year = 2018,
        month = oct,
       volume = {866},
       number = {2},
          eid = {110},
        pages = {110},
          doi = {10.3847/1538-4357/aadadd},
archivePrefix = {arXiv},
       eprint = {1808.06613},
 primaryClass = {astro-ph.EP},
       adsurl = {https://ui.adsabs.harvard.edu/abs/2018ApJ...866..110D}
}

@ARTICLE{2003A&A...399..297W,
       author = {{Woitke}, P. and {Helling}, Ch.},
        title = "{Dust in brown dwarfs. II. The coupled problem of dust formation and sedimentation}",
      journal = {\aap},
         year = 2003,
        month = feb,
       volume = {399},
        pages = {297-313},
          doi = {10.1051/0004-6361:20021734},
       adsurl = {https://ui.adsabs.harvard.edu/abs/2003A&A...399..297W}
}

@BOOK{2020apfs.book.....A,
       author = {{Armitage}, Philip J.},
        title = "{Astrophysics of planet formation, Second Edition}",
         year = 2020,
         publisher= {Cambridge University Press},
       adsurl = {https://ui.adsabs.harvard.edu/abs/2020apfs.book.....A}
}

@ARTICLE{2025MNRAS.543.4198C,
       author = {{Cordwell}, Amelia J. and {Ziampras}, Alexandros and {Brown}, Joshua J. and {Rafikov}, Roman R.},
        title = "{How two-dimensional are planet─disc interactions? I. Locally isothermal discs}",
      journal = {\mnras},
         year = 2025,
        month = nov,
       volume = {543},
       number = {4},
        pages = {4198-4217},
          doi = {10.1093/mnras/staf1674},
archivePrefix = {arXiv},
       eprint = {2509.04282},
 primaryClass = {astro-ph.EP},
       adsurl = {https://ui.adsabs.harvard.edu/abs/2025MNRAS.543.4198C}
}

@ARTICLE{2020MNRAS.493.5892W,
       author = {{Wafflard-Fernandez}, Gaylor and {Baruteau}, Cl{\'e}ment},
        title = "{Intermittent planet migration and the formation of multiple dust rings and gaps in protoplanetary discs}",
      journal = {\mnras},
         year = 2020,
        month = apr,
       volume = {493},
       number = {4},
        pages = {5892-5912},
          doi = {10.1093/mnras/staa379},
archivePrefix = {arXiv},
       eprint = {2002.02280},
 primaryClass = {astro-ph.EP},
       adsurl = {https://ui.adsabs.harvard.edu/abs/2020MNRAS.493.5892W}
}

@ARTICLE{2015A&A...573A...9P,
       author = {{Pinilla}, P. and {de Juan Ovelar}, M. and {Ataiee}, S. and {Benisty}, M. and {Birnstiel}, T. and {van Dishoeck}, E.~F. and {Min}, M.},
        title = "{Gas and dust structures in protoplanetary disks hosting multiple planets}",
      journal = {\aap},
         year = 2015,
        month = jan,
       volume = {573},
          eid = {A9},
        pages = {A9},
          doi = {10.1051/0004-6361/201424679},
archivePrefix = {arXiv},
       eprint = {1410.5963},
 primaryClass = {astro-ph.EP},
       adsurl = {https://ui.adsabs.harvard.edu/abs/2015A&A...573A...9P}
}

@ARTICLE{2017A&A...608A..92D,
       author = {{Dr{\k{a}}{\.z}kowska}, J. and {Alibert}, Y.},
        title = "{Planetesimal formation starts at the snow line}",
      journal = {\aap},
         year = 2017,
        month = dec,
       volume = {608},
          eid = {A92},
        pages = {A92},
          doi = {10.1051/0004-6361/201731491},
archivePrefix = {arXiv},
       eprint = {1710.00009},
 primaryClass = {astro-ph.EP},
       adsurl = {https://ui.adsabs.harvard.edu/abs/2017A&A...608A..92D}
}

@ARTICLE{testi2022,
       author = {{Testi}, L. and {Natta}, A. and {Manara}, C.~F. and {de Gregorio Monsalvo}, I. and {Lodato}, G. and {Lopez}, C. and {Muzic}, K. and {Pascucci}, I. and {Sanchis}, E. and {Miranda}, A. Santamaria and {Scholz}, A. and {De Simone}, M. and {Williams}, J.~P.},
        title = "{The protoplanetary disk population in the {\ensuremath{\rho}}-Ophiuchi region L1688 and the time evolution of Class II YSOs}",
      journal = {\aap},
         year = 2022,
        month = jul,
       volume = {663},
          eid = {A98},
        pages = {A98},
          doi = {10.1051/0004-6361/202141380},
archivePrefix = {arXiv},
       eprint = {2201.04079},
 primaryClass = {astro-ph.SR},
       adsurl = {https://ui.adsabs.harvard.edu/abs/2022A&A...663A..98T}
}

@ARTICLE{barnabo2022,
       author = {{Bernab{\`o}}, Lia Marta and {Turrini}, Diego and {Testi}, Leonardo and {Marzari}, Francesco and {Polychroni}, Danai},
        title = "{Dust Resurgence in Protoplanetary Disks Due to Planetesimal-Planet Interactions}",
      journal = {\apjl},
         year = 2022,
        month = mar,
       volume = {927},
       number = {2},
          eid = {L22},
        pages = {L22},
          doi = {10.3847/2041-8213/ac574e},
       adsurl = {https://ui.adsabs.harvard.edu/abs/2022ApJ...927L..22B}
}

@ARTICLE{turrini2019,
       author = {{Turrini}, D. and {Marzari}, F. and {Polychroni}, D. and {Testi}, L.},
        title = "{Dust-to-gas Ratio Resurgence in Circumstellar Disks Due to the Formation of Giant Planets: The Case of HD 163296}",
      journal = {\apj},
         year = 2019,
        month = may,
       volume = {877},
       number = {1},
          eid = {50},
        pages = {50},
          doi = {10.3847/1538-4357/ab18f5},
archivePrefix = {arXiv},
       eprint = {1802.04361},
 primaryClass = {astro-ph.EP},
       adsurl = {https://ui.adsabs.harvard.edu/abs/2019ApJ...877...50T}
}

@ARTICLE{zhu2022,
       author = {{Zhu}, Wei},
        title = "{The Intrinsic Multiplicity Distribution of Exoplanets Revealed from the Radial Velocity Method}",
      journal = {\aj},
         year = 2022,
        month = jul,
       volume = {164},
       number = {1},
          eid = {5},
        pages = {5},
          doi = {10.3847/1538-3881/ac6f59},
archivePrefix = {arXiv},
       eprint = {2201.03782},
 primaryClass = {astro-ph.EP},
       adsurl = {https://ui.adsabs.harvard.edu/abs/2022AJ....164....5Z}
}

@ARTICLE{knutson2014,
       author = {{Knutson}, Heather A. and {Fulton}, Benjamin J. and {Montet}, Benjamin T. and {Kao}, Melodie and {Ngo}, Henry and {Howard}, Andrew W. and {Crepp}, Justin R. and {Hinkley}, Sasha and {Bakos}, Gaspar {\'A}. and {Batygin}, Konstantin and {Johnson}, John Asher and {Morton}, Timothy D. and {Muirhead}, Philip S.},
        title = "{Friends of Hot Jupiters. I. A Radial Velocity Search for Massive, Long-period Companions to Close-in Gas Giant Planets}",
      journal = {\apj},
         year = 2014,
        month = apr,
       volume = {785},
       number = {2},
          eid = {126},
        pages = {126},
          doi = {10.1088/0004-637X/785/2/126},
archivePrefix = {arXiv},
       eprint = {1312.2954},
 primaryClass = {astro-ph.EP},
       adsurl = {https://ui.adsabs.harvard.edu/abs/2014ApJ...785..126K}
}

@ARTICLE{matsumura2021,
       author = {{Matsumura}, Soko and {Brasser}, Ramon and {Ida}, Shigeru},
        title = "{N-body simulations of planet formation via pebble accretion. II. How various giant planets form}",
      journal = {\aap},
         year = 2021,
        month = jun,
       volume = {650},
          eid = {A116},
        pages = {A116},
          doi = {10.1051/0004-6361/202039210},
archivePrefix = {arXiv},
       eprint = {2104.07271},
 primaryClass = {astro-ph.EP},
       adsurl = {https://ui.adsabs.harvard.edu/abs/2021A&A...650A.116M}
}

@ARTICLE{ida2013,
       author = {{Ida}, S. and {Lin}, D.~N.~C. and {Nagasawa}, M.},
        title = "{Toward a Deterministic Model of Planetary Formation. VII. Eccentricity Distribution of Gas Giants}",
      journal = {\apj},
         year = 2013,
        month = sep,
       volume = {775},
       number = {1},
          eid = {42},
        pages = {42},
          doi = {10.1088/0004-637X/775/1/42},
archivePrefix = {arXiv},
       eprint = {1307.6450},
 primaryClass = {astro-ph.EP},
       adsurl = {https://ui.adsabs.harvard.edu/abs/2013ApJ...775...42I}
}

@ARTICLE{charnoz2011,
       author = {{Charnoz}, S{\'e}bastien and {Fouchet}, Laure and {Aleon}, J{\'e}r{\^o}me and {Moreira}, Manuel},
        title = "{Three-dimensional Lagrangian Turbulent Diffusion of Dust Grains in a Protoplanetary Disk: Method and First Applications}",
      journal = {\apj},
         year = 2011,
        month = aug,
       volume = {737},
       number = {1},
          eid = {33},
        pages = {33},
          doi = {10.1088/0004-637X/737/1/33},
archivePrefix = {arXiv},
       eprint = {1105.3440},
 primaryClass = {astro-ph.EP},
       adsurl = {https://ui.adsabs.harvard.edu/abs/2011ApJ...737...33C}
}

@ARTICLE{garate2020,
       author = {{G{\'a}rate}, Mat{\'\i}as and {Birnstiel}, Til and {Dr{\k{a}}{\.z}kowska}, Joanna and {Stammler}, Sebastian Markus},
        title = "{Gas accretion damped by dust back-reaction at the snow line}",
      journal = {\aap},
         year = 2020,
        month = mar,
       volume = {635},
          eid = {A149},
        pages = {A149},
          doi = {10.1051/0004-6361/201936067},
archivePrefix = {arXiv},
       eprint = {1906.07708},
 primaryClass = {astro-ph.EP},
       adsurl = {https://ui.adsabs.harvard.edu/abs/2020A&A...635A.149G}
}

@ARTICLE{mignone2007,
       author = {{Mignone}, A. and {Bodo}, G. and {Massaglia}, S. and {Matsakos}, T. and {Tesileanu}, O. and {Zanni}, C. and {Ferrari}, A.},
        title = "{PLUTO: A Numerical Code for Computational Astrophysics}",
      journal = {\apjs},
         year = 2007,
        month = may,
       volume = {170},
       number = {1},
        pages = {228-242},
          doi = {10.1086/513316},
archivePrefix = {arXiv},
       eprint = {astro-ph/0701854},
 primaryClass = {astro-ph},
       adsurl = {https://ui.adsabs.harvard.edu/abs/2007ApJS..170..228M}
}

@ARTICLE{rosotti2020,
       author = {{Rosotti}, Giovanni P. and {Teague}, Richard and {Dullemond}, Cornelis and {Booth}, Richard A. and {Clarke}, Cathie J.},
        title = "{The efficiency of dust trapping in ringed protoplanetary discs}",
      journal = {\mnras},
         year = 2020,
        month = jun,
       volume = {495},
       number = {1},
        pages = {173-181},
          doi = {10.1093/mnras/staa1170},
archivePrefix = {arXiv},
       eprint = {2004.11394},
 primaryClass = {astro-ph.EP},
       adsurl = {https://ui.adsabs.harvard.edu/abs/2020MNRAS.495..173R}
}

@ARTICLE{pinilla2012,
       author = {{Pinilla}, P. and {Benisty}, M. and {Birnstiel}, T.},
        title = "{Ring shaped dust accumulation in transition disks}",
      journal = {\aap},
         year = 2012,
        month = sep,
       volume = {545},
          eid = {A81},
        pages = {A81},
          doi = {10.1051/0004-6361/201219315},
archivePrefix = {arXiv},
       eprint = {1207.6485},
 primaryClass = {astro-ph.EP},
       adsurl = {https://ui.adsabs.harvard.edu/abs/2012A&A...545A..81P}
}

@INPROCEEDINGS{lissauer1993,
       author = {{Lissauer}, Jack J. and {Stewart}, Glen R.},
        title = "{Growth of Planets from Planetesimals}",
    booktitle = {Protostars and Planets III},
         year = 1993,
       editor = {{Levy}, Eugene H. and {Lunine}, Jonathan I.},
        month = jan,
        pages = {1061},
       adsurl = {https://ui.adsabs.harvard.edu/abs/1993prpl.conf.1061L}
}

@ARTICLE{marzari_angelo2019,
       author = {{Marzari}, Francesco and {D'Angelo}, Gennaro and {Picogna}, Giovanni},
        title = "{Circumstellar Dust Distribution in Systems with Two Planets in Resonance}",
      journal = {\aj},
         year = 2019,
        month = feb,
       volume = {157},
       number = {2},
          eid = {45},
        pages = {45},
          doi = {10.3847/1538-3881/aaf3b6},
archivePrefix = {arXiv},
       eprint = {1812.07698},
 primaryClass = {astro-ph.EP},
       adsurl = {https://ui.adsabs.harvard.edu/abs/2019AJ....157...45M}
}

@ARTICLE{isella2016,
       author = {{Isella}, Andrea and {Guidi}, Greta and {Testi}, Leonardo and {Liu}, Shangfei and {Li}, Hui and {Li}, Shengtai and {Weaver}, Erik and {Boehler}, Yann and {Carperter}, John M. and {De Gregorio-Monsalvo}, Itziar and {Manara}, Carlo F. and {Natta}, Antonella and {P{\'e}rez}, Laura M. and {Ricci}, Luca and {Sargent}, Anneila and {Tazzari}, Marco and {Turner}, Neal},
        title = "{Ringed Structures of the HD 163296 Protoplanetary Disk Revealed by ALMA}",
      journal = {\prl},
         year = 2016,
        month = dec,
       volume = {117},
       number = {25},
          eid = {251101},
        pages = {251101},
          doi = {10.1103/PhysRevLett.117.251101},
       adsurl = {https://ui.adsabs.harvard.edu/abs/2016PhRvL.117y1101I}
}

@ARTICLE{weidenschilling1977,
       author = {{Weidenschilling}, S.~J.},
        title = "{Aerodynamics of solid bodies in the solar nebula.}",
      journal = {\mnras},
         year = 1977,
        month = jul,
       volume = {180},
        pages = {57-70},
          doi = {10.1093/mnras/180.2.57},
       adsurl = {https://ui.adsabs.harvard.edu/abs/1977MNRAS.180...57W}
}

@ARTICLE{linpapa1986,
       author = {{Lin}, D.~N.~C. and {Papaloizou}, J.},
        title = "{On the Tidal Interaction between Protoplanets and the Primordial Solar Nebula. II. Self-Consistent Nonlinear Interaction}",
      journal = {\apj},
         year = 1986,
        month = aug,
       volume = {307},
        pages = {395},
          doi = {10.1086/164426},
       adsurl = {https://ui.adsabs.harvard.edu/abs/1986ApJ...307..395L}
}

@ARTICLE{dzyu2010,
       author = {{Dzyurkevich}, N. and {Flock}, M. and {Turner}, N.~J. and {Klahr}, H. and {Henning}, Th.},
        title = "{Trapping solids at the inner edge of the dead zone: 3-D global MHD simulations}",
      journal = {\aap},
         year = 2010,
        month = jun,
       volume = {515},
          eid = {A70},
        pages = {A70},
          doi = {10.1051/0004-6361/200912834},
archivePrefix = {arXiv},
       eprint = {1002.2521},
 primaryClass = {astro-ph.SR},
       adsurl = {https://ui.adsabs.harvard.edu/abs/2010A&A...515A..70D}
}

@ARTICLE{riols_MHD,
       author = {{Riols}, A. and {Lesur}, G.},
        title = "{Spontaneous ring formation in wind-emitting accretion discs}",
      journal = {\aap},
         year = 2019,
        month = may,
       volume = {625},
          eid = {A108},
        pages = {A108},
          doi = {10.1051/0004-6361/201834813},
archivePrefix = {arXiv},
       eprint = {1904.07910},
 primaryClass = {astro-ph.EP},
       adsurl = {https://ui.adsabs.harvard.edu/abs/2019A&A...625A.108R}
}

@ARTICLE{flock2015,
       author = {{Flock}, M. and {Ruge}, J.~P. and {Dzyurkevich}, N. and {Henning}, Th. and {Klahr}, H. and {Wolf}, S.},
        title = "{Gaps, rings, and non-axisymmetric structures in protoplanetary disks. From simulations to ALMA observations}",
      journal = {\aap},
         year = 2015,
        month = feb,
       volume = {574},
          eid = {A68},
        pages = {A68},
          doi = {10.1051/0004-6361/201424693},
archivePrefix = {arXiv},
       eprint = {1411.2736},
 primaryClass = {astro-ph.EP},
       adsurl = {https://ui.adsabs.harvard.edu/abs/2015A&A...574A..68F}
}

@ARTICLE{zhang2015,
       author = {{Zhang}, Ke and {Blake}, Geoffrey A. and {Bergin}, Edwin A.},
        title = "{Evidence of Fast Pebble Growth Near Condensation Fronts in the HL Tau Protoplanetary Disk}",
      journal = {\apjl},
         year = 2015,
        month = jun,
       volume = {806},
       number = {1},
          eid = {L7},
        pages = {L7},
          doi = {10.1088/2041-8205/806/1/L7},
archivePrefix = {arXiv},
       eprint = {1505.00882},
 primaryClass = {astro-ph.EP},
       adsurl = {https://ui.adsabs.harvard.edu/abs/2015ApJ...806L...7Z}
}

@ARTICLE{dong_fung2017,
       author = {{Dong}, Ruobing and {Fung}, Jeffrey},
        title = "{What is the Mass of a Gap-opening Planet?}",
      journal = {\apj},
         year = 2017,
        month = feb,
       volume = {835},
       number = {2},
          eid = {146},
        pages = {146},
          doi = {10.3847/1538-4357/835/2/146},
archivePrefix = {arXiv},
       eprint = {1612.04821},
 primaryClass = {astro-ph.EP},
       adsurl = {https://ui.adsabs.harvard.edu/abs/2017ApJ...835..146D}
}

@ARTICLE{crida2006,
       author = {{Crida}, A. and {Morbidelli}, A. and {Masset}, F.},
        title = "{On the width and shape of gaps in protoplanetary disks}",
      journal = {\icarus},
         year = 2006,
        month = apr,
       volume = {181},
       number = {2},
        pages = {587-604},
          doi = {10.1016/j.icarus.2005.10.007},
archivePrefix = {arXiv},
       eprint = {astro-ph/0511082},
 primaryClass = {astro-ph},
       adsurl = {https://ui.adsabs.harvard.edu/abs/2006Icar..181..587C}
}

@ARTICLE{ALMA2015,
       author = {{ALMA Partnership} and {Brogan}, C.~L. and {P{\'e}rez}, L.~M. and {Hunter}, T.~R. and {Dent}, W.~R.~F. and {Hales}, A.~S. and {Hills}, R.~E. and {Corder}, S. and {Fomalont}, E.~B. and {Vlahakis}, C. and {Asaki}, Y. and {Barkats}, D. and {Hirota}, A. and {Hodge}, J.~A. and {Impellizzeri}, C.~M.~V. and {Kneissl}, R. and {Liuzzo}, E. and {Lucas}, R. and {Marcelino}, N. and {Matsushita}, S. and {Nakanishi}, K. and {Phillips}, N. and {Richards}, A.~M.~S. and {Toledo}, I. and {Aladro}, R. and {Broguiere}, D. and {Cortes}, J.~R. and {Cortes}, P.~C. and {Espada}, D. and {Galarza}, F. and {Garcia-Appadoo}, D. and {Guzman-Ramirez}, L. and {Humphreys}, E.~M. and {Jung}, T. and {Kameno}, S. and {Laing}, R.~A. and {Leon}, S. and {Marconi}, G. and {Mignano}, A. and {Nikolic}, B. and {Nyman}, L. -A. and {Radiszcz}, M. and {Remijan}, A. and {Rod{\'o}n}, J.~A. and {Sawada}, T. and {Takahashi}, S. and {Tilanus}, R.~P.~J. and {Vila Vilaro}, B. and {Watson}, L.~C. and {Wiklind}, T. and {Akiyama}, E. and {Chapillon}, E. and {de Gregorio-Monsalvo}, I. and {Di Francesco}, J. and {Gueth}, F. and {Kawamura}, A. and {Lee}, C. -F. and {Nguyen Luong}, Q. and {Mangum}, J. and {Pietu}, V. and {Sanhueza}, P. and {Saigo}, K. and {Takakuwa}, S. and {Ubach}, C. and {van Kempen}, T. and {Wootten}, A. and {Castro-Carrizo}, A. and {Francke}, H. and {Gallardo}, J. and {Garcia}, J. and {Gonzalez}, S. and {Hill}, T. and {Kaminski}, T. and {Kurono}, Y. and {Liu}, H. -Y. and {Lopez}, C. and {Morales}, F. and {Plarre}, K. and {Schieven}, G. and {Testi}, L. and {Videla}, L. and {Villard}, E. and {Andreani}, P. and {Hibbard}, J.~E. and {Tatematsu}, K.},
        title = "{The 2014 ALMA Long Baseline Campaign: First Results from High Angular Resolution Observations toward the HL Tau Region}",
      journal = {\apjl},
         year = 2015,
        month = jul,
       volume = {808},
       number = {1},
          eid = {L3},
        pages = {L3},
          doi = {10.1088/2041-8205/808/1/L3},
archivePrefix = {arXiv},
       eprint = {1503.02649},
 primaryClass = {astro-ph.SR},
       adsurl = {https://ui.adsabs.harvard.edu/abs/2015ApJ...808L...3A}
}

@ARTICLE{avenhaus_sphere_2018,
       author = {{Avenhaus}, Henning and {Quanz}, Sascha P. and {Garufi}, Antonio and {Perez}, Sebastian and {Casassus}, Simon and {Pinte}, Christophe and {Bertrang}, Gesa H. -M. and {Caceres}, Claudio and {Benisty}, Myriam and {Dominik}, Carsten},
        title = "{Disks around T Tauri Stars with SPHERE (DARTTS-S). I. SPHERE/IRDIS Polarimetric Imaging of Eight Prominent T Tauri Disks}",
      journal = {\apj},
         year = 2018,
        month = aug,
       volume = {863},
       number = {1},
          eid = {44},
        pages = {44},
          doi = {10.3847/1538-4357/aab846},
archivePrefix = {arXiv},
       eprint = {1803.10882},
 primaryClass = {astro-ph.SR},
       adsurl = {https://ui.adsabs.harvard.edu/abs/2018ApJ...863...44A}
}

@ARTICLE{isella2018ApJ,
       author = {{Isella}, Andrea and {Huang}, Jane and {Andrews}, Sean M. and {Dullemond}, Cornelis P. and {Birnstiel}, Tilman and {Zhang}, Shangjia and {Zhu}, Zhaohuan and {Guzm{\'a}n}, Viviana V. and {P{\'e}rez}, Laura M. and {Bai}, Xue-Ning and {Benisty}, Myriam and {Carpenter}, John M. and {Ricci}, Luca and {Wilner}, David J.},
        title = "{The Disk Substructures at High Angular Resolution Project (DSHARP). IX. A High-definition Study of the HD 163296 Planet-forming Disk}",
      journal = {\apjl},
         year = 2018,
        month = dec,
       volume = {869},
       number = {2},
          eid = {L49},
        pages = {L49},
          doi = {10.3847/2041-8213/aaf747},
archivePrefix = {arXiv},
       eprint = {1812.04047},
 primaryClass = {astro-ph.SR},
       adsurl = {https://ui.adsabs.harvard.edu/abs/2018ApJ...869L..49I}
}

@ARTICLE{2018A&A...616A..47T,
       author = {{Thun}, Daniel and {Kley}, Wilhelm},
        title = "{Migration of planets in circumbinary discs}",
      journal = {\aap},
         year = 2018,
        month = aug,
       volume = {616},
          eid = {A47},
        pages = {A47},
          doi = {10.1051/0004-6361/201832804},
archivePrefix = {arXiv},
       eprint = {1806.00314},
 primaryClass = {astro-ph.EP},
       adsurl = {https://ui.adsabs.harvard.edu/abs/2018A&A...616A..47T}
}

@software{2012ascl.soft02015D,
       author = {{Dullemond}, C.~P. and {Juhasz}, A. and {Pohl}, A. and {Sereshti}, F. and {Shetty}, R. and {Peters}, T. and {Commercon}, B. and {Flock}, M.},
        title = "{RADMC-3D: A multi-purpose radiative transfer tool}",
 howpublished = {Astrophysics Source Code Library, record ascl:1202.015},
         year = 2012,
        month = feb,
          eid = {ascl:1202.015},
archivePrefix = {ascl},
       eprint = {1202.015},
       adsurl = {https://ui.adsabs.harvard.edu/abs/2012ascl.soft02015D}
}

@ARTICLE{2019MNRAS.486..304B,
       author = {{Baruteau}, Cl{\'e}ment and {Barraza}, Marcelo and {P{\'e}rez}, Sebasti{\'a}n and {Casassus}, Simon and {Dong}, Ruobing and {Lyra}, Wladimir and {Marino}, Sebasti{\'a}n and {Christiaens}, Valentin and {Zhu}, Zhaohuan and {Carmona}, Andr{\'e}s and {Debras}, Florian and {Alarcon}, Felipe},
        title = "{Dust traps in the protoplanetary disc MWC 758: two vortices produced by two giant planets?}",
      journal = {\mnras},
         year = 2019,
        month = jun,
       volume = {486},
       number = {1},
        pages = {304-319},
          doi = {10.1093/mnras/stz802},
archivePrefix = {arXiv},
       eprint = {1903.06537},
 primaryClass = {astro-ph.EP},
       adsurl = {https://ui.adsabs.harvard.edu/abs/2019MNRAS.486..304B}
}

@ARTICLE{2003ARA&A..41..241D,
       author = {{Draine}, B.~T.},
        title = "{Interstellar Dust Grains}",
      journal = {\araa},
         year = 2003,
        month = jan,
       volume = {41},
        pages = {241-289},
          doi = {10.1146/annurev.astro.41.011802.094840},
archivePrefix = {arXiv},
       eprint = {astro-ph/0304489},
 primaryClass = {astro-ph},
       adsurl = {https://ui.adsabs.harvard.edu/abs/2003ARA&A..41..241D}
}

@ARTICLE{2007Icar..192..588Y,
       author = {{Youdin}, Andrew N. and {Lithwick}, Yoram},
        title = "{Particle stirring in turbulent gas disks: Including orbital oscillations}",
      journal = {\icarus},
         year = 2007,
        month = dec,
       volume = {192},
       number = {2},
        pages = {588-604},
          doi = {10.1016/j.icarus.2007.07.012},
archivePrefix = {arXiv},
       eprint = {0707.2975},
 primaryClass = {astro-ph},
       adsurl = {https://ui.adsabs.harvard.edu/abs/2007Icar..192..588Y}
}

@ARTICLE{2025A&A...703A.180P,
       author = {{Parizek}, Facundo and {Planes}, Mar{\'\i}a Bel{\'e}n and {Mill{\'a}n}, Emmanuel Nicol{\'a}s and {Parisi}, M. Gabriela and {Bringa}, Eduardo Marcial},
        title = "{Dust collisions in protoplanetary disks: From monodisperse to bidisperse grain aggregates}",
      journal = {\aap},
         year = 2025,
        month = nov,
       volume = {703},
          eid = {A180},
        pages = {A180},
          doi = {10.1051/0004-6361/202556240},
       adsurl = {https://ui.adsabs.harvard.edu/abs/2025A&A...703A.180P}
}

@ARTICLE{2024ARA&A..62..157B,
       author = {{Birnstiel}, Tilman},
        title = "{Dust Growth and Evolution in Protoplanetary Disks}",
      journal = {\araa},
         year = 2024,
        month = sep,
       volume = {62},
       number = {1},
        pages = {157-202},
          doi = {10.1146/annurev-astro-071221-052705},
archivePrefix = {arXiv},
       eprint = {2312.13287},
 primaryClass = {astro-ph.EP},
       adsurl = {https://ui.adsabs.harvard.edu/abs/2024ARA&A..62..157B}
}

@ARTICLE{2018A&A...616A.116P,
       author = {{Picogna}, Giovanni and {Stoll}, Moritz H.~R. and {Kley}, Wilhelm},
        title = "{Particle accretion onto planets in discs with hydrodynamic turbulence}",
      journal = {\aap},
         year = 2018,
        month = aug,
       volume = {616},
          eid = {A116},
        pages = {A116},
          doi = {10.1051/0004-6361/201732523},
archivePrefix = {arXiv},
       eprint = {1803.08730},
 primaryClass = {astro-ph.EP},
       adsurl = {https://ui.adsabs.harvard.edu/abs/2018A&A...616A.116P}
}

@ARTICLE{2025A&A...703A.270R,
       author = {{Roatti}, V. and {Picogna}, G. and {Marzari}, F.},
        title = "{Dust distribution in circumstellar disks harboring multi-planet systems: I. Sub-thermal mass planets}",
      journal = {\aap},
         year = 2025,
        month = nov,
       volume = {703},
          eid = {A270},
        pages = {A270},
          doi = {10.1051/0004-6361/202556463},
archivePrefix = {arXiv},
       eprint = {2511.06050},
 primaryClass = {astro-ph.EP},
       adsurl = {https://ui.adsabs.harvard.edu/abs/2025A&A...703A.270R}
}

@ARTICLE{2014ApJ...785..122Z,
       author = {{Zhu}, Zhaohuan and {Stone}, James M. and {Rafikov}, Roman R. and {Bai}, Xue-ning},
        title = "{Particle Concentration at Planet-induced Gap Edges and Vortices. I. Inviscid Three-dimensional Hydro Disks}",
      journal = {\apj},
         year = 2014,
        month = apr,
       volume = {785},
       number = {2},
          eid = {122},
        pages = {122},
          doi = {10.1088/0004-637X/785/2/122},
archivePrefix = {arXiv},
       eprint = {1308.0648},
 primaryClass = {astro-ph.EP},
       adsurl = {https://ui.adsabs.harvard.edu/abs/2014ApJ...785..122Z}
}

@ARTICLE{2009A&A...493.1125L,
       author = {{Lyra}, W. and {Johansen}, A. and {Klahr}, H. and {Piskunov}, N.},
        title = "{Standing on the shoulders of giants. Trojan Earths and vortex trapping in low mass self-gravitating protoplanetary disks of gas and solids}",
      journal = {\aap},
         year = 2009,
        month = jan,
       volume = {493},
       number = {3},
        pages = {1125-1139},
          doi = {10.1051/0004-6361:200810797},
archivePrefix = {arXiv},
       eprint = {0810.3192},
 primaryClass = {astro-ph},
       adsurl = {https://ui.adsabs.harvard.edu/abs/2009A&A...493.1125L}
}

@ARTICLE{2018A&A...612A..30B,
       author = {{Bitsch}, Bertram and {Morbidelli}, Alessandro and {Johansen}, Anders and {Lega}, Elena and {Lambrechts}, Michiel and {Crida}, Aur{\'e}lien},
        title = "{Pebble-isolation mass: Scaling law and implications for the formation of super-Earths and gas giants}",
      journal = {\aap},
         year = 2018,
        month = apr,
       volume = {612},
          eid = {A30},
        pages = {A30},
          doi = {10.1051/0004-6361/201731931},
archivePrefix = {arXiv},
       eprint = {1801.02341},
 primaryClass = {astro-ph.EP},
       adsurl = {https://ui.adsabs.harvard.edu/abs/2018A&A...612A..30B}
}

@ARTICLE{2014A&A...572A..35L,
       author = {{Lambrechts}, M. and {Johansen}, A. and {Morbidelli}, A.},
        title = "{Separating gas-giant and ice-giant planets by halting pebble accretion}",
      journal = {\aap},
         year = 2014,
        month = dec,
       volume = {572},
          eid = {A35},
        pages = {A35},
          doi = {10.1051/0004-6361/201423814},
archivePrefix = {arXiv},
       eprint = {1408.6087},
 primaryClass = {astro-ph.EP},
       adsurl = {https://ui.adsabs.harvard.edu/abs/2014A&A...572A..35L}
}

@software{2021ascl.soft04010D,
       author = {{Dominik}, Carsten and {Min}, Michiel and {Tazaki}, Ryo},
        title = "{OpTool: Command-line driven tool for creating complex dust opacities}",
 howpublished = {Astrophysics Source Code Library, record ascl:2104.010},
         year = 2021,
        month = apr,
          eid = {ascl:2104.010},
archivePrefix = {ascl},
       eprint = {2104.010},
       adsurl = {https://ui.adsabs.harvard.edu/abs/2021ascl.soft04010D}
}

@ARTICLE{1973A&A....24..337S,
       author = {{Shakura}, N.~I. and {Sunyaev}, R.~A.},
        title = "{Black holes in binary systems. Observational appearance.}",
      journal = {\aap},
         year = 1973,
        month = jan,
       volume = {24},
        pages = {337-355},
       adsurl = {https://ui.adsabs.harvard.edu/abs/1973A&A....24..337S}
}

@ARTICLE{2012A&A...541A.123M,
       author = {{M{\"u}ller}, T.~W.~A. and {Kley}, W. and {Meru}, F.},
        title = "{Treating gravity in thin-disk simulations}",
      journal = {\aap},
         year = 2012,
        month = may,
       volume = {541},
          eid = {A123},
        pages = {A123},
          doi = {10.1051/0004-6361/201118737},
archivePrefix = {arXiv},
       eprint = {1203.1413},
 primaryClass = {astro-ph.EP},
       adsurl = {https://ui.adsabs.harvard.edu/abs/2012A&A...541A.123M}
}

@ARTICLE{2006MNRAS.370..529D,
       author = {{de Val-Borro}, M. and {Edgar}, R.~G. and {Artymowicz}, P. and {Ciecielag}, P. and {Cresswell}, P. and {D'Angelo}, G. and {Delgado-Donate}, E.~J. and {Dirksen}, G. and {Fromang}, S. and {Gawryszczak}, A. and {Klahr}, H. and {Kley}, W. and {Lyra}, W. and {Masset}, F. and {Mellema}, G. and {Nelson}, R.~P. and {Paardekooper}, S. -J. and {Peplinski}, A. and {Pierens}, A. and {Plewa}, T. and {Rice}, K. and {Sch{\"a}fer}, C. and {Speith}, R.},
        title = "{A comparative study of disc-planet interaction}",
      journal = {\mnras},
         year = 2006,
        month = aug,
       volume = {370},
       number = {2},
        pages = {529-558},
          doi = {10.1111/j.1365-2966.2006.10488.x},
archivePrefix = {arXiv},
       eprint = {astro-ph/0605237},
 primaryClass = {astro-ph},
       adsurl = {https://ui.adsabs.harvard.edu/abs/2006MNRAS.370..529D}
}

@ARTICLE{2025A&A...700A.190R,
       author = {{Ruzza}, A. and {Lodato}, G. and {Rosotti}, G.~P. and {Armitage}, P.~J.},
        title = "{DBNets2.0: Simulation-based inference for planet-induced dust substructures in protoplanetary discs}",
      journal = {\aap},
         year = 2025,
        month = aug,
       volume = {700},
          eid = {A190},
        pages = {A190},
          doi = {10.1051/0004-6361/202554401},
archivePrefix = {arXiv},
       eprint = {2506.11200},
 primaryClass = {astro-ph.EP},
       adsurl = {https://ui.adsabs.harvard.edu/abs/2025A&A...700A.190R}
}

\begin{appendix}
\onecolumn
    \section{Dust continuum emission}
\begin{figure*}[h!]
     \centering
     \includegraphics[width = \linewidth, height=0.7\textheight,keepaspectratio]{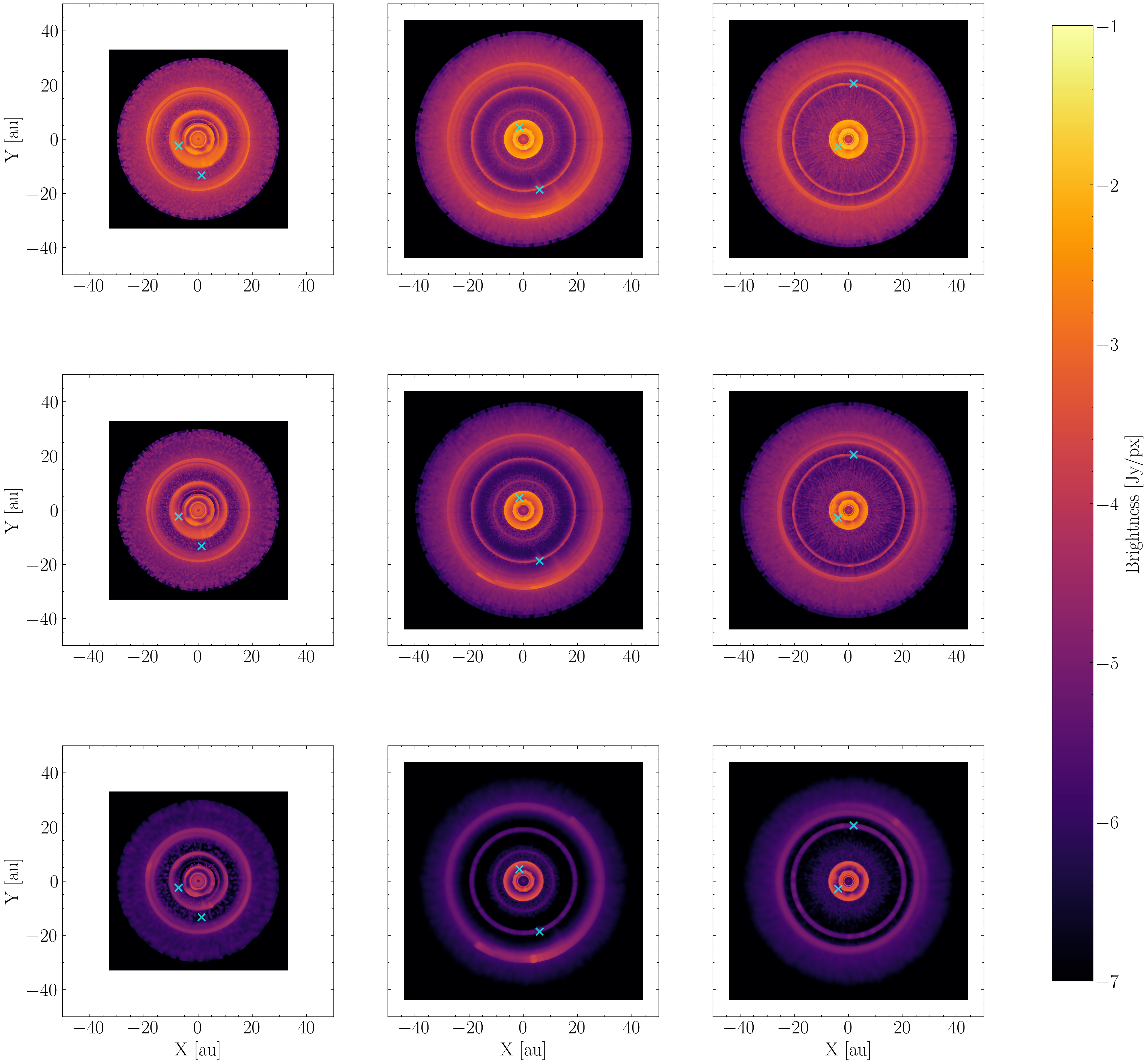}
     \caption{Continuum emission raw maps in Band 7 (top row), Band 6 (middle row) and Band 3 (bottom row), for the models with two Jupiter-mass planets in close orbits (left column), wide orbits (middle column) and for a Jupiter-Saturn pair (right column). Crosses indicate the position of the planets.}
      \label{fig:raw_brightness}
 \end{figure*}

        \cifig{fig:raw_brightness} shows the raw emission maps (before beam convolution) for the different multi-planet systems simulated in Band 7 (0.85 mm), Band 6 (1.3 mm) and Band 3 (3 mm). In all three configurations, the intensity of the dust emission has a maximum close to the star, which makes it hard to notice the gap created by the inner planet. The outer dust gap is more visible, in particular at longer wavelengths. A big difference between the close-orbit and the wide-orbit configurations is that the first one shows two gaps at the location of the planets, while in the second case a dust ring forms corresponding to the orbit of the outer planet. This could be an artifact due to particles that would have been accreted by the planet, or could represent a large number of dust particles trapped in the horseshoe region because the planets orbiting farther out did not have enough time to "clear" their orbits with respect to the closer configuration.\\* 

\section{Dust vortex}
\begin{figure*}[h!]
     \centering
     \includegraphics[width = \linewidth,keepaspectratio]{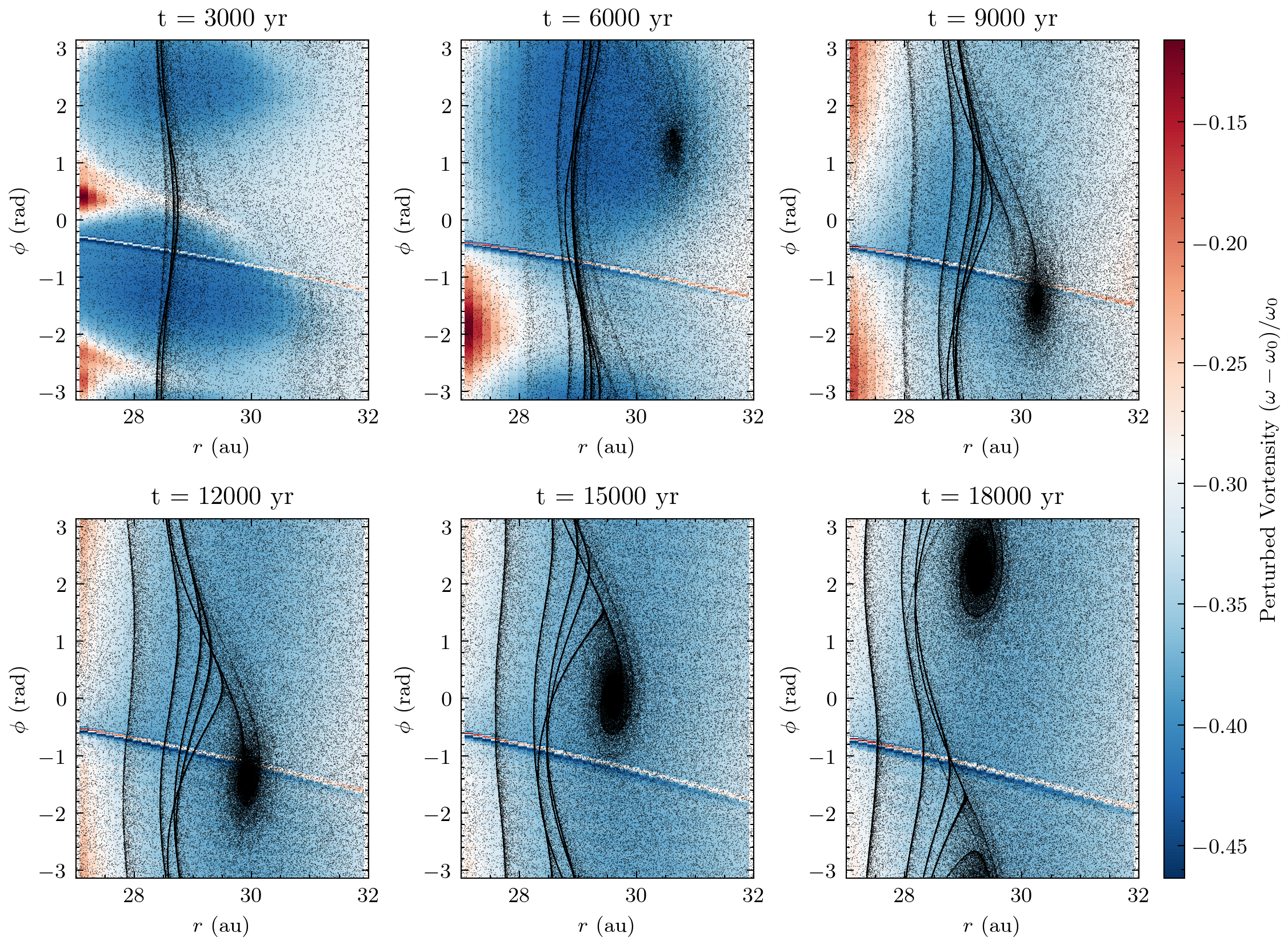}
     \caption{Perturbed vortensity, $(\omega - \omega_0)/\omega_0$ (color scale), overlaid with the distribution of dust particles of all sizes (black dots) at different times for the model with two Jupiter-mass planets on wide orbits. The outer planet is fixed at $\phi = 0$.}
      \label{fig:vortensity}
 \end{figure*}

We have shown in \cifig{fig:snapshot} the presence of an azimuthal asymmetry in the dust distribution, which originates from an initial perturbation of the gas vortensity. In 2D, the vortensity, $\omega$, is defined as the ratio of the curl of the velocity field to the gas surface density, $\Sigma$. \cifig{fig:vortensity} presents the time evolution of the perturbed vortensity, $(\omega - \omega_0)/\omega_0$, overlaid with the distribution of dust particles of all sizes. Although the initial vortensity perturbation gradually dissipates, a dusty vortex forms at the same location and grows with time. The resulting dust overdensity produces the peak in continuum emission shown in \cifig{radmc}.

\twocolumn
\end{appendix}
\end{document}